# The Effect of Physical Assumptions on the Calculation of Microwave Background Anisotropies

Wayne Hu[1], Douglas Scott[1], Naoshi Sugiyama[1,2] & Martin White[1]

[1] *Center for Particle Astrophysics
and Departments of Astronomy and Physics,
University of California, Berkeley, CA 94720-7304*

[2] *Department of Physics, Faculty of Science,
University of Tokyo, Tokyo, 113, Japan*

As the data on cosmic microwave background anisotropies improve and potential cosmological applications are realized, it will be increasingly important for theoretical calculations to be as accurate as possible. All modern calculations for inflationary-inspired fluctuations involve the numerical solution of coupled Boltzmann equations. There are many assumptions and choices to be made when carrying out such calculations. Here we go through each assumption in turn, pointing out the best selections to make in each case, and the level of inaccuracy expected through incorrect choice. For example, neglecting the effects of neutrinos or polarization has a 10% effect. Varying input parameters such as radiation temperature and helium fraction can have smaller, but noticeable effects. We also discuss a few issues which are more numerical, such as $k$-range and smoothing. Some short-cut methods for obtaining the anisotropy spectrum are also investigated, for example free-streaming and tilt approximations; generally none of these are adequate at the few % level. At the level of 1% it is important to consider somewhat baroque effects, such as helium recombination and even minimal amounts of reionization. At smaller angular scales there are secondary and higher-order effects which will ultimately have to be considered. Extracting information from the subsidiary acoustic peaks and the damping region of the anisotropy spectrum will be an extremely challenging problem. However, given the real prospect of measuring just such information on the sky, it will be important to meet this challenge. In principle it will be possible to extract rather detailed information about reionization history, neutrino contribution, helium abundance, non-power-law initial conditions etc.

e-mail: hu@pac2.berkeley.edu
dscott@roma.berkeley.edu
sugiyama@pac3.berkeley.edu
white@pac1.berkeley.edu

*Le bon Dieu est dans le détail.*
   — *Gustave Flaubert*

## 1. INTRODUCTION

It now seems clear that anisotropies in the Cosmic Microwave Background (CMB) have been detected over a wide range of angular scales (see [1], for a recent review). We are in a time of rapid experimental progress, and there is realistic hope of being able to measure fundamental cosmological parameters from the shape of the power spectrum of anisotropies. It has been shown that there are combinations of parameters which are 'degenerate', i.e. quite different cosmological models can give very similar anisotropies [2]. However, these degeneracies are not exact [3,4,5,6], and to understand the precise sensitivity to specific parameters, it is necessary to carry out calculations which are as accurate as possible. In fact, satellite experiments which are now being considered may be able to map most of the sky down to a fraction of a degree, thereby measuring a range of cosmological quantities at once. However the size of the variations which need to be measured to extract full cosmological information is significantly smaller than the accuracy with which theoretical predictions have been routinely given in the past (cf. tables in [7] for a comparison of results).

In order to keep pace with the expected progress in experiments, we need to examine the robustness of the theoretical calculations. In this paper, we have set about studying the various assumptions which go into these calculations. For most of the issues, there is a very definite correct choice in order to obtain the most accurate results, although in some circumstances it may be better to compromise accuracy in favour of computational speed. Although many of these effects are already included in modern state-of-the-art Boltzmann codes, in general none of them have been fully discussed in the literature. Moreover, there are other issues where the assumptions are perhaps more subtle or implicit, and only become apparent in a systematic investigation.

The outline of the paper is as follows. As discussed in §II, the power spectrum of CMB anisotropies is usually denoted by the quantities $C_\ell$ (squares of the amplitudes in a spherical harmonic decomposition $\ell$ [8]) for each multipole and calculated by solving the radiative transport or Boltzmann equations. As a rule of thumb, we will suppose that we are interested in calculating these $C_\ell$'s to an accuracy of at least 1% out to arcminute scales, since this is roughly what will be required by experiments currently being proposed. We begin by considering the effect on anisotropies of background quantities in §III, i.e. the matter content, expansion rate, and recombination history of the universe. In §IV, we expand the scope to include sources in linear perturbation theory which are subdominant and often omitted in calculations. Of course, it should be stressed that on sub-degree angular scales there are sources of anisotropy which are likely to be $\gtrsim 1\%$ of the primary anisotropies which are our focus in this paper. We also survey such 'secondary' or 'non-linear' effects in §IV. However understanding the issues involved in obtaining primary $C_\ell$'s accurate to 1% is still a worthwhile exercise and a necessary first step in calculating small scale anisotropies. Finally in §V, we discuss the accuracy of numerical approximations commonly employed to make the calculation tractable.

Our general approach in this paper is to understand what physical effects and input assumptions give measurable changes in the predictions of the $C_\ell$'s for a fiducial model, taken to be a cold dark matter (CDM) model with adiabatic initial conditions in an $\Omega_0 = 1$ universe. Of course many of our results will extend to other models. Having established this goal, it is a simple matter to reverse the argument to determine the physical processes and parameters that we could *measure* by obtaining accurate $C_\ell$'s from the sky.

## II. THE BOLTZMANN EQUATIONS

The standard way to calculate the present-day spectrum of anisotropies is to write down the Fourier transform of the linearized Boltzmann equation for each cosmological fluid (dark matter, baryons, photons, and neutrinos), with all of the relevant physical couplings between the fluids retained. These equations have been presented many times in the literature, and since they require the precise definition of many different quantities, we avoid their repetition here. Instead we refer the interested reader to the seminal papers of [9,10,11,12,13,8]. More recent treatments can be found in [14,15] (following the approach of [16]) and [17].

These equations can be solved semi-analytically for either standard recombination, using the tight-coupling approximation for the photons and baryons (e.g. [18,19,20,21,22,23,24, 25,26,3]) or for reionization, using the weak coupling approximation (e.g. [27,28,29,14,30, 15]). Methods for both cases are further developed and refined in [4,5]. The accuracy of the calculations is limited by the ability to fit or calculate the time-dependence of the gravitational potential. Although such techniques are extremely useful for understanding the underlying physical mechanisms, and are likely to remain useful for some time to come, ultimately the full numerical solution of the coupled Boltzmann equations will be required in order to obtain the most accurate $C_\ell$'s.

Detailed numerical calculations of the Boltzmann equations have been carried out for many different models, e.g. the pioneering work of [9,10,13,8,31,32], and the more recent treatments by [33,34,35,7,36,37,6] amongst others. The basic method is due to the work of [13,8] (see also [38, 39]). By integrating over frequencies, the full Boltzmann equation for the photons reduces to an evolution equation for the brightness or temperature fluctuation. This approach is valid in linear perturbation theory since spectral distortions are only introduced at second order, provided there is no exotic source of energy injection into the CMB. The angular dependence of each $k$-mode of the radiation can be expanded in a series of Legendre polynomials [10], which reduces the equations to an infinite hierarchy of coupled ordinary differential equations. [This development was a vast, and physically-sensible, improvement over earlier treatments which performed calculations using discrete $\mu$-modes.] At very high redshift it is sufficient to use the tight–coupling approximation for photons and baryons. In this approximation (described more in §V.B), the Compton scattering rate is sufficiently rapid that anisotropies in the photon distribution cannot be generated and the modes with $\ell \geq 2$ are exponentially damped. Only the density ($\ell = 0$) and velocity ($\ell = 1$) perturbations are kept in the photon and baryon components. This is often referred to as the perfect fluid approximation, since the two components can be described by density and pressure perturbations alone. Viscosity in the fluid can be treated perturbatively in this limit and leads to a damping of small scale anisotropies. As the Universe begins to recombine, the moments $\ell \geq 2$ are no longer exponentially damped and the full hierarchy must be evolved.

The outcome of the numerical evolution is a catalogue of solutions of the Boltzmann equations for the angular moments of each $k$-mode [38]. These can then be combined using an initial weighting of the $k$-modes (i.e. an initial power spectrum) to obtain the $C_\ell$'s. Providing that *enough* modes are used and *all* relevant sources are included, the calculation can then be carried out to an accuracy of $10^{-5}$, due to the perturbation expansion and the measured level of CMB fluctuations. We shall be concerned in this paper not so much with the application of this method of solution as with the physical inputs and approximations which are made to render this program tractable. However in §V, we will comment on some subtle issues of this method of solution which can cause inaccuracies to creep in.

In order to show the quantitative size of the effects we discuss, we choose a specific cosmological model around which to vary our different assumptions and other physical inputs. For definiteness, we will take $\Omega_0 = 1$, $\Omega_B = 0.05$, $h = 0.5$, $n = 1$, $T/S = 0$, i.e. the standard Cold Dark Matter model (sCDM) with adiabatic Gaussian initial conditions and no tensor component. We will discuss more specific assumptions and refine our model as we progress. The anisotropy spectrum for this model is shown in Fig. 1. This is plotted in the usual form of $\ell(\ell+1)C_\ell$ vs. $\ell$, which gives the power per logarithmic interval in $\ell$.

## III. THE BACKGROUND COSMOLOGY

### A. Photon Temperature and Neutrinos

One of the most basic input parameters is the temperature of the microwave background today. By altering the radiation content of the universe this affects the epochs of last scattering, matter-radiation equality and baryon-photon equality, and hence the anisotropies in the CMB. A simple question then arises: what is the effect of using a slightly different value? In Fig. 2 we show the ratio of the angular power spectrum, $\ell(\ell+1)C_\ell$ vs $\ell$, for three temperatures $T_{\gamma 0} = 3$ K, 2.7 K, and 2.73 K, to that for the current best value of 2.726 K [40]. This shows that the $\pm 0.005$ K uncertainty in the FIRAS measurement of the temperature leads to a < 1% change in the $C_\ell$'s.

Other massless species such as neutrinos affect the redshift of matter radiation equality and hence the growth rate of perturbations. Specifically, any increase in the relativistic content prevents the growth of perturbations inside the horizon. This leads to decay in the gravitational potential, which affects the photons through gravitational redshift [41]. In Fig. 3(a,b), we show the effect of varying the number of neutrinos from 2 to 4 which changes $C_\ell$ on the order of $\pm 10\%$ near the first peak. The decay of the potential drives the acoustic oscillations, resulting in larger anisotropies. Notice that the deviation with radiative content has a distinct and symmetrical signature near the first peak that is potentially measurable and competitive with nucleosynthesis bounds. At smaller scales, changes in the angular location of the peaks through the horizon and sound horizon at last scattering, as well as additional effects in the damping regime, lead to a more complicated, oscillatory structure in the ratio. Nonetheless, the CMB is an efficient probe of the epoch of matter radiation equality [42], and accurate measurement of the $C_\ell$'s can be used to constrain the relativistic matter content of the universe [43].

Since the case of massive neutrinos can effectively be described as a universe whose relativistic content decreases as the neutrinos become non-relativistic, these arguments show that the CMB can in principle yield information on the mass of the neutrino (see [7,44,45] for a full treatment). The qualitatively new feature is that the angle that the horizon subtends when the neutrino becomes non-relativistic ($\ell \simeq 500$ for $m_\nu$ of a few eV) will be imprinted into the CMB as the transition between a CDM spectrum and the higher relativistic content case.

### B. Baryons and the Hubble Constant

Once the present photon temperature is fixed, the baryon content of the universe governs the evolution of temperature perturbations through the sound speed in the photon-baryon fluid [9]. Raising the baryon content, i.e. $\Omega_B h^2$, changes the balance of pressure and gravity, leading to more fluctuation growth. The Hubble constant has two other effects. It changes the distance scale of the universe, which has little effect on anisotropies, and it changes the matter-radiation ratio. As discussed above in the context of neutrinos, the matter-radiation ratio affects the growth of perturbations and controls the gravitational redshift contributions. The two effects in $h$ are competitive if $\Omega_B \simeq 0.05$. Fixing $h = 0.5$, the change in $C_\ell$'s from standard CDM to $\Omega_B h^2 = 0.015$ is 6% at $\ell \simeq 200$, whereas fixing $\Omega_B h^2 = 0.0125$, the change to $h = 0.55$ is 10% (see Fig. 4,5). Bond *et al.* [2] point out that the two parameters are difficult to separate. With a satellite however, it may be feasible to obtain cosmic variance limited $C_\ell$'s up to the first peak, in which case the parameters are distinguishable. In any case, beyond the first peak the degeneracy is completely broken since pressure and gravity affect odd and even peaks differently [3,4].

At extremely small scales ($k \gtrsim 100\,h\,\mathrm{Mpc}^{-1}$), i.e. below the sound horizon of the thermal baryon fluid, baryonic pressure must be included [17]. This does not however affect anisotropies for any conceivable experiment or model since they arise from much larger scales at last scattering. However, it is possible that it could become important in reionized scenarios with extensive heating, since this pressure prevents Jeans instability for the baryons.

We must also set a value for the fraction of baryons contained in helium (He). Since helium recombines earlier than hydrogen, this has a substantial effect on temperature anisotropies by changing the free electron density at last scattering (i.e. there is less than one free electron per baryon, $n_e \propto (1 - Y_p)\Omega_B h^2$). Our fiducial model has a primordial helium mass fraction $Y_P = 0.23$. We show in Fig. 6 the differences obtained by adopting the values $Y_P = 0.20$ and $Y_P = 0.26$. As can be seen, it is necessary to include Helium as a neutral component of the universe in order to obtain accurate $C_\ell$'s. We shall now discuss to what extent the recombination history of both hydrogen and helium affect anisotropies.

### C. Recombination

#### 1. Physical Considerations

The process of recombination would proceed via the Saha equation (see e.g. [46]), except that recombinations to the ground state are inhibited by the recombination process itself [47]. Thus recombination is controlled by the population of the first excited state, and the physical processes which either populate or depopulate it in the expanding Universe. This problem was first worked out in detail by [48] and at about the same time by [49], and by many authors since [50,51].

Solving the coupled ionization and matter temperature equations gives the evolution of the ionized fraction $x_e(z) \equiv n_e/n_H$ and the visibility function $g(z) \equiv e^{-\tau} d\tau/dz$, for Thomson scattering optical depth $\tau$. The quantity $g(z)dz$ is the fraction of the radiation that was last scattered in a redshift interval $dz$. There are in general two effects that the ionization history has on the $C_\ell$'s. The visibility function determines the epoch at which fluctuations from the tight coupling regime are frozen in. The ionization fraction $x_e(z)$ determines the breakdown of tight coupling, i.e the photon diffusion length, which is responsible for the damping of anisotropies. Thus the detailed description of the ionization in the tails of the visibility function is important: two ionization histories with the same maximum and width for their visibility functions will not in general lead to the same $C_\ell$'s.

This is most clearly seen in the consideration of helium recombination. One might naively expect it to have a negligible effect on the $C_\ell$'s because Helium recombines while the radiation and matter are still very tightly coupled, at $z \simeq 2500$ for HeII and $z \simeq 6000$ for HeIII. However the diffusion damping length grows continuously and is sensitive to the full thermal history. We find that inclusion of helium recombination affects the 2nd, 3rd and 4th peaks at the 0.2%, 0.4% and 1% levels, as shown in Fig. 7. This is in good agreement with analytic estimates which interpret the effect as a decrease in photon diffusion during the epoch that helium was ionized. Hence it *is* important to follow the recombination of the helium in order to obtain accurate $C_\ell$'s at the percent level. [On the other hand, the effects of D, $^3$He, Li etc. are entirely negligible until the $10^{-5}$ level.] Fig. 7 demonstrates that even seemingly minor effects on the ionization history can yield changes in the $C_\ell$'s at the percent level. Note that because of atomic collisions, even after helium recombination, helium atoms are tightly coupled to the hydrogen through collisions. Since they contribute to the inertia of the photon-baryon fluid, helium atoms should therefore be kept in the baryon evolution equations.

However, we found that following HeIII → HeII was not entirely necessary, and furthermore that simple use of the Saha equation for Helium is as good as following the helium atoms more fully. Treating Helium in isolation one finds that the HeII to HeI recombination is slower than the Saha equation predicts [52,53,54], (though note that in principle more levels are important than those followed in this reference, and the energy gap between $n = 2$ and the continuum is only $\simeq 3$eV for HeII, so the recombination would happen at $z \simeq 1000$ in contrast to the results of [52,53,54]). However, there is a trace of neutral Hydrogen present even at redshifts $z \simeq 2500$, which can absorb the Helium Ly$\alpha$ photons and prevent every Helium recombination photon from ionizing another Helium [55]. We find that by $z = 2500$ the mean free time for a neutral Hydrogen atom to capture a Helium Ly$\alpha$ photon is many orders of magnitude smaller than the Hubble time. The extra energy imparted to the electron is rapidly shared with the plasma, which is strongly enough coupled to the photons at this redshift that the matter temperature is unaffected. Thus direct recombination to the ground state *is* possible for Helium (in contrast with the case for Hydrogen) and the ionized fraction follows the Saha value to a good approximation. We show the recombination of the stages of helium and hydrogen together in Fig. 8 (specifically plotted for the standard CDM cosmological parameters).

It is worth noting that the framework in which the recombination calculation is done does not allow for the inclusion of spatial information, i.e. the effect of the inhomogeneities on the recombination process. However, this is a second-order effect. In fact, it is just like the Vishniac effect (§IV.D.2), except at high redshift when the baryons and photons are still tightly coupled. Thus we expect it to be extremely small.

### 2. Refinements of Hydrogen Recombination

Since the recombination process is so crucial to both the calculation and interpretation of CMB anisotropies, and because much of the work on recombination considers new effects in isolation, we shall consider a range of approximations and refinements to the hydrogen recombination process. The first level of approximation is to use the Saha equation in an expanding universe, which is shown by the dashed curve in Fig. 9. This approximation has long been realized to be inadequate and most recent work has followed, for example, the equations of [50]. However, it has also been realised that following recombination to even higher accuracy can be important [56] while imposing little computational burden compared with evolving the temperature anisotropies.

Further refinements to recombination will have much smaller, but still potentially important effects. One variation involves following the kinetic temperature ($T_{\mathrm{mat}}$) of the electrons independently [57,48]. It turns out that this has an effect only on the very late-time tail of the visibility function, since the matter and radiation remain coupled until $z \sim 200$ (see Fig. 10). The effect on the $C_\ell$'s is on the order 0.1% and is thus negligible. On the other hand, an exotic scenario involving energy injection can change the matter temperature significantly and can in principle be constrained through the damping tail of anisotropies.

Within these assumptions, the full differential equation for the ionization fraction contains many terms in principle, although most can be ignored. For example the extra terms (due to stimulated recombination etc.) included in [50] (or equivalently [58]) have negligible effect, as can be seen in their comparison with the results of [48]. Although collisions are important for keeping the various matter species in kinetic equilibrium, the collisional rates for excitation and ionization are negligible at the relevant densities and temperatures [52,53]. Radiative transfer effects in the Ly $\alpha$ line are also negligible [59,60,51]. Furthermore it was shown recently [61] that the standard assumption of a quasi-static solution for the Ly $\alpha$ profile is good to at least 1 part in $10^5$. It has been suggested [62] that the $2p$ and $2s$ states in Hydrogen might not be expected to be in statistical equilibrium. This would increase $x_e$ by a small amount at low $z$, although with little effect on the visibility function and the $C_\ell$'s. However, it is clear that the $2p$ and $2s$ states would be strongly coupled by the electric fields of nearby atoms or ions at the rapid collision rate during recombination [52,53]. Hence this non-equilibrium effect does not actually occur.

There is also a potential extra stimulated emission effect, pointed out by [63], caused by the non-equilibrium between the excited states and the continuum. (These authors perform a classical calculation of the recombination coefficients, and do not include the Gaunt factors. Among other things these factors take into account the asymptotic state of the electron, which at low energies is affected by the nuclear charge even at large distances). This would appear to have a measurable effect on the $C_\ell$'s through affecting the low $z$ tail of the visibility function. However, it would seem that such out of equilibrium considerations can only be studied in detail by following the populations of the levels in a hydrogen atom in detail (Sasselov & Scott, in preparation). We expect such effects to be potentially important at the 1 or 2 % level in the $C_\ell$'s, but are unlikely to be larger. We are led to the conclusion that at the present the major real source of ambiguity comes from the assumed values for the recombination rates, so we now discuss this in more detail.

The single most important effect in obtaining accurate power spectra is using the most accurate recombination coefficient $\alpha$. The specific coefficient which is needed is the sum of all direct recombinations excluding those to the ground state, often denoted $\alpha_B$ ('case B' recombination is when the Lyman lines are optically thick). Although the rate to an individual level has $\alpha \propto T_{\rm mat}^{-1/2}$, this is not at all true for the sum over many transitions (although [48,50] and other authors have used this approximation). A better approximation is that $\alpha_B \propto T_{\rm mat}^{-3/4}$ (e.g. [64]), but this is only good for $T_{\rm mat} \sim 10^4{\rm K}$, which is a much higher temperature than we are interested in.

There are better rates available in the literature, which are accurate over a wide range of temperatures and even have fitted functional forms, e.g. [65]. Recently there has been more work on obtaining the most accurate recombination rates. Currently the best are the values tabulated by [66], giving $\alpha_B$ accurate to the 4th significant figure for temperatures all the way down to 10 K in steps of 0.2 in $\log T_{\rm mat}$. There is also a fitting function given by [67], which is accurate to $< 0.2\%$:

$$\alpha_B = 10^{-13} \frac{a\,t^b}{1 + c\,t^d} \ {\rm cm}^3{\rm s}^{-1},$$

where $a = 4.309$, $b = -0.6166$, $c = 0.6703$, $d = 0.5300$ and $t = T_{\rm mat}/10^4{\rm K}$. Although these rates are calculated in the limit of zero density, it is apparent from the table in [66] that for cosmological recombination the effect of the electron density is never more than about 2%, and significantly less at the lower redshifts which are most relevant. We show in Fig. 7 the effect that including the correct recombination rate as opposed to a $T^{1/2}$ scaling has on the $C_\ell$'s. Notice that it can be several percent over a wide range of scales. We find that the approximation to $\alpha_B$ of [17] or those of [68] or [65] give essentially the same $C_\ell$'s as these more accurate rates.

It is also worth updating other quantities relevant for calculating the recombination. For example, the two-photon rate from the 2s to 1s states has been re-calculated recently (e.g. [69]), and is $\Lambda_{2s \to 1s} = 8.22458 {\rm s}^{-1}$, including the reduced mass correction. This improvement however has an entirely negligible effect on the $C_\ell$'s.

None of the standard references for recombination take into account the earlier period of radiation domination. This is important especially for low $\Omega_0 h^2$ models, since the equality epoch does not occur very much earlier than the recombination epoch. We found that assuming the complete matter-dominated background evolution resulted in $x_e$ being systematically a few percent lower (depending on the exact parameters), decreasing the damping. This is because assuming $z_{\rm eq} \to \infty$ lowers the Hubble rate, meaning more recombination for a given redshift. For an $\Omega_0 = 1$ model the exact relation $H(z) = H_0 a^{-2}(a + a_{\rm eq})^{1/2}$ can be easily used.

### D. Physical Constant & Other Uncertainties

It is obviously also important to use the most accurate available physical constants and parameters. One fundamental limitation to accuracy is set by the uncertainty in the gravitational constant $G$, which will affect the overall timescale. However, even this quantity is known to an accuracy of $\simeq 10^{-4}$ (Particle Data Group 1994), and all other relevant numbers are known much more precisely.

Compton scattering of protons is reduced by $(m_e/m_p)^2$ compared with the electrons, and is thus negligible. Relativistic corrections for Compton scattering off electrons will be $\mathcal{O}(10^{-5})$, even at redshifts $\sim 10^4$. There are however ambiguities $\mathcal{O}(m_e/m_p)$ in going from $\Omega_B$ to $n_e$, although again this is a small effect.

Other particle physics effects could potentially influence the calculations. For example there could be non-negligible interactions or velocities in the dark matter, chemical potential, decays or other extra physics in the neutrino sector, strong primordial magnetic fields, etc. However, it seems contrived for such effects to play a role at the percent level, without totally altering the picture of structure formation.

## IV. SUBDOMINANT ANISOTROPY SOURCES

### A. Neutrino Fluctuations

Anisotropies may offer the first, albeit crude and indirect, probe of *fluctuations* in the neutrino background through their gravitational feedback on the CMB (see also §III.A). Evolving the temperature perturbations in the neutrinos involves another 'infinite' hierarchy of $\ell$-modes that must be solved. As a first approximation, one might consider the neutrinos to be a perfect fluid and truncate the hierarchy for $\ell \geq 2$, i.e. consider only the density and velocity (pressure) perturbations. This approximation neglects the free-streaming damping and anisotropic stress contributions of the neutrinos. Both feed back into the evolution of the perturbations in the metric, and thus the temperature anisotropies in the photons [41]. In Fig. 11, we show that this induces an error of about 10%. Since physically this dependence is coming entirely from the neutrino monopole, dipole and quadrupole, it is evident that there is no need to keep the hierarchy for $\ell > 2$. However, free streaming monotonically transfers power to high $\ell$ (a certain scale subtends an angle on the sky which decreases with the distance between source and observer) so naive truncation of the hierarchy produces an artificial reflection of power to low $\ell$'s from the last $\ell$ kept. This can be avoided by an appropriate choice of boundary conditions for the maximum $\ell$-mode of the neutrinos, using the analytic free streaming solution for the neutrinos, i.e. the recurrence relations among the spherical Bessel functions $j_\ell$ may be used to modify the coupling of the last $\ell$ mode kept, once it enters its oscillatory phase. This allows truncation of the $\ell$ hierarchy at very low $\ell$. While the phases of the last $\ell$ modes kept are not accurately preserved, there is no reflection of power to lower $\ell$ modes, thus the effect of neutrino anisotropy on the $C_\ell$ is correctly accounted for. Furthermore, this boundary condition is also applicable to the photon-Boltzmann equation during the oscillatory phase. Using this boundary condition, we do not have to solve for the very high $\ell$ modes which we are not interested in.

However, for completeness we also show in Fig. 12 the present day anisotropy spectrum of the massless neutrinos. Although this spectrum is a clear prediction of standard hot Big Bang model, we do *not* expected it to be tested in the near future! Despite the fact that there are more than 300 background neutrinos per cm$^3$, they are extremely low energy and therefore essentially impossible to detect. And measuring fluctuations in the background at the level of $10^{-5}$ will be somewhat harder! The specific calculation shown in the figure is for the parameters of standard CDM, with three species of light neutrinos. The amplitude of the fluctuations, although plotted dimensionlessly, is exactly the same as for the photons. [This is true for $\Delta T_\nu/T_\nu$, while $\Delta T_\nu$, measured in e.g. $\mu$K, would be $(4/11)^{1/3}$ lower in each species]. At small $\ell$ the anisotropies come from the familiar Sachs-Wolfe effect from potential fluctuations. For the largest $\ell$'s, there is an extra contribution from the integrated Sachs-Wolfe effect, since potentials decay after they come inside the horizon during radiation domination. This effect is the same for all sufficiently small scales, with an amplitude which is $\simeq 15$ times greater than at large scales. [This is the same size as the extra kick that the photons get, as in §III.A, except that they have scattering and oscillations as well.] The characteristic scale of the step is $\ell \sim 500$, which corresponds to the angle subtended by the horizon size at the epoch of matter-radiation equality. The small wiggles are a real effect that can be attributed to the oscillating density of the photon-baryon fluid, which is most important before equality. Note that the roughly flat $\ell(\ell+1)C_\ell$ does not continue to infinite $\ell$ (thus maintaining finite total power), since there will be a cut-off corresponding to the diffusion-scale at neutrino decoupling, at $\ell \sim 10^8$.

## B. Polarization

Polarization is generated at the last scattering surface by Compton scattering of anisotropic radiation because the Thomson cross section depends on angle as $|\epsilon_f \cdot \epsilon_i|^2$, where $\epsilon_f$ and $\epsilon_i$ are the final and initial polarization vectors respectively. Furthermore, polarization feeds back into anisotropies. Averaging over incident and summing over final polarizations leads to an angular dependence: $1 + \cos^2 \theta$. Since the scattering of linearly polarized radiation will in general have a different angular dependence than this, the scattering term in the Boltzmann equation for temperature perturbations will be modified by polarization. (A pedagogical treatment has recently been given by [70]). More specifically, the quadrupole moment of the temperature distribution leads to linear polarization in the microwave background (e.g. [71,72]) and vice versa [13]. The precise level of the temperature anisotropies therefore is not recovered by neglecting polarization (as has recently been emphasized by Bond and Steinhardt). In fact, polarization leads to an increased damping of anisotropies [72].

The easiest way to see how polarization affects temperature perturbations is to (artificially) extend the tight coupling approximation through last scattering. Recall that in the tight coupling approximation only density and velocity fluctuations are present (higher terms are damped) and the matter and radiation behave as a fluid described by density and pressure. Specifically any temperature quadrupole is strongly damped. Since polarization is a source of quadrupole anisotropies, this represents a breakdown of tight coupling due to the generation of viscosity [72]. The corresponding effective increase in the photon 'diffusion scale' leads to an increase in the damping angle by 4-5% and an increasingly large error in $C_\ell$ with $\ell$, due to the near exponential behavior of the damping, as shown in Fig. 13. In low $\Omega_B$ models, as is required by nucleosynthesis, the diffusion scale is already large at last scattering and the additional effect of polarization leads to smaller anisotropies even near the first peak.

The power spectrum of polarization anisotropies was shown in the second panel of Fig. 1, for the sCDM model. Note that in reionized scenarios the polarization can extend to much larger scales due to the increase in the horizon scale at last scattering. It is also worth pointing out that circular polarization is not generated through scattering, unless there are large coherent magnetic fields or other exotic phenomena. Finally, as a computational note, if only the temperature feedback effect is of interest, the polarization evolution equations may be dropped after last scattering with no loss of accuracy.

## C. Gravity Waves

In addition to the scalar modes with which the previous discussion has been involved, there is the possibility that inflation excites tensor (i.e. gravity wave) perturbations as well [73]. Most of the points mentioned above apply also to the calculation of tensor perturbations. [Note that vector perturbations only have decaying modes, and so are unimportant if the fluctuations were generated in the early universe]. Early work on tensors and the CMB was performed by [74,75,76]. There exist several semi-analytic approximations of varying accuracy, the most recent and accurate being due to [77]. To calculate the tensor spectrum numerically one uses the formalism of [78] as first worked out in detail by [36]. This leads to another set of Boltzmann equations, independent of those for the scalars, which follow the temperature and polarization anisotropies of the tensors. The final result is then $C_\ell^{\text{tot}} = C_\ell^{(S)} + C_\ell^{(T)}$ where the relative normalization of the tensor and scalar components depends on the details of the perturbation generation scenario.

The tensor modes evolve completely independently from the scalar modes in linear theory, i.e. the resulting temperature patterns are uncorrelated. Polarization and temperature fluctuations due to gravity waves on the other hand are correlated in an analogous but distinct manner compared with the scalar case [79,80]. Note also that since the gravity wave $C_\ell$'s drop off at the horizon-size at last-scattering, it is not critical to accurately follow recombination, polarization, high $k$-modes, etc. for the calculation of the *total* temperature anisotropy. For example including polarization changes the $C_\ell$ by only 4% at $\ell \simeq 100$, where they are already a small fraction of $C_\ell^{\text{tot}}$. In Fig. 14, we plot the tensor contribution $C_\ell^{(T)}$, for a model with the parameters of standard CDM.

### D. Secondary Anisotropies

#### 1. Reionization

Reionization in a CDM scenario is unlikely to occur at redshift $z \gtrsim 50$ [81,82]. Since in a fully ionized $x_e = 1$ universe (assuming the sCDM parameters), the optical depth $\tau$ does not reach unity until $z \simeq 100$, reionization is probably not a dominant effect. However, it can still make a contribution to the $C_\ell$'s. In general, reionization has two effects on the CMB (see e.g. [83,27,84]). Under the horizon scale at last scattering, it damps power in the primary anisotropies as $e^{-2\tau}$ (see Figs. 13, 14). It is important to realize that even for minimal reionization there may be a measurable effect. We know from null results on the Gunn-Peterson test [85], that the Universe was largely ionized out to at least $z = 4$–$5$ [86,87]. For sCDM this corresponds to $\tau \simeq 0.01$, and hence will affect the $C_\ell$'s at the 2% level (see Fig. 15). Of course this variation would be lost in the cosmic variance at small $\ell$. However, it could still result in erroneous conclusions about other parameters. It should be remembered that this is absolutely the smallest possible reionization effect, and thus the CMB can be used to probe the epoch of reionization even if it occurred at low redshift.

Difficulties arise if reionization is patchy, we hope that if $\tau$ is large, many regions are averaged over. Inhomogeneous reionization can also induce secondary anisotropies on the CMB through preferential scattering. As well as creating patchy reionization damping of the primary signal, secondary anisotropies can also be generated. Small fluctuations in ionization can be expected to behave like the Vishniac effect described below (see [14]). Large fluctuations are essentially identical to the peculiar velocity cluster Sunyaev-Zel'dovich effect.

In addition to damping under the horizon scale at last scattering, secondary anisotropies are generated on the new last scattering surface by Doppler shifts from scattering. Since both damping processes and fluctuation growth depend on the epoch of last scattering, it is *not* sufficient to parameterize the ionization history by the total optical depth. In fact, reionization allows not only free parameters but a free function $x_e(z)$. However, for the anisotropy spectrum, most of the effect can be described by two parameters, the optical depth and the epoch of last scattering. As long as the universe is dominated by cold dark matter, so that fluctuation growth does not depend on the thermal history when Compton drag was important, any model which predicts the same late time visibility function will yield the same anisotropy (unlike anisotropies originating at recombination). In Fig. 16, we display two models with the same $\tau$ but with different visibility functions to illustrate this point. Specifically the two models have $\tau \simeq 15\%$, with $x_e = 1$ out to $z = 30$ and $x_e = 0.5$ out to $z = 48$.

#### 2. The Vishniac Effect

In linear theory, fluctuations on scales smaller than the width of the visibility function ($\simeq 5'$ in standard recombination) are strongly damped. Ref. [88] showed that there are second order effects for which this is not true, due to the geometry of the cancellation process. One such second order contribution, coming from a product of velocities and densities, dominates over all others [14,15] and is known as the Vishniac effect ([28], see also [89,29,90]). The contributions are entirely negligible for standard recombination scenarios since second order terms are necessarily small and baryon velocities are suppressed by Silk damping. However, if the universe was reionized at some late epoch, infall into CDM wells and the growth of fluctuations in the intervening period allows for a significant contribution. Indeed, the Vishniac contribution is so highly peaked toward late times that Gunn-Peterson constraints *require* some contribution from this effect. For a fully ionized universe, the Vishniac effect is larger than the primary signal calculated here at $\ell \simeq 3,000$ [91]. The effect itself peaks at $\ell \simeq 7,000$ and has $\ell(\ell + 1)C_\ell \sim 0.025$ [units of $6C_2$] at the peak for a fully ionized, COBE normalized model and decreases only gradually as the ionization redshift is lowered. Note that since the Vishniac effect is second order, its amplitude relative to primary anisotropies depends on the normalization employed.

#### 3. The Rees-Sciama Effect

Even in an $\Omega_0 = 1$ universe, the non-linear growth of structure will cause time variations in the gravitational potential which lead to anisotropies via gravitational redshifts [41]. Two effects can be identified: one is due to the growth of the potential during the photon transit time across the fluctuation [92] and the other is due to a spatially varying potential crossing the line of sight. By ray tracing techniques through N-body simulations of CDM, the two effects have been shown to be comparable in magnitude and contribute to $\Delta T/T$ at the level of a few $\times 10^{-7}$ (i.e. $0.01 - 0.1\%$ in $C_\ell$) at degree scales ($\ell \sim 200$) [93]. It is possible that the effect increases significantly toward arcminute scales in some models as a result of non-linear evolution [94]. Recently however calculations based on N-body results have shown that this effect only dominates the primary fluctuation at $\ell \gtrsim 5000$ for CDM [95].

### 4. The Cluster Sunyaev-Zel'dovich Effect

As pointed out by Sunyaev and Zel'dovich [96], clusters can also induce anisotropies on the CMB from Compton scattering off electrons in the hot cluster medium. These hot electrons transfer energy to the microwave background, leading to temperature anisotropies *and* spectral distortions in the CMB. Thus the temperature fluctuation will not only have an angular but also a frequency dependence, unlike primary sources of anisotropies. A large amount of work has been done to try to estimate the fluctuations caused by S-Z fluctuations (e.g. [97,98,99,100,101,102,103, 104,105,106]) with varying conclusions.

Recently [107] have employed an empirically based model for clusters. They find that in the Rayleigh-Jeans regime, where the Sunyaev-Zel'dovich effect leads to a constant brightness decrement, the anisotropy at arcminutes is on the order $\Delta T/T \lesssim 10^{-7}$. So we expect an effect on the $C_\ell$'s of order $\lesssim 0.01 - 0.1\%$ at $\ell \gtrsim 1000$. Moreover, the signal is in large part due to bright and easily identifiable clusters. If such known clusters are removed from the sample, the anisotropy drops to an entirely negligible level.

Clusters can also provide anisotropies since their peculiar velocity leads to a Doppler shift of the scattered photons. This process leads to no spectral distortions to first order and is of order $\Delta T/T \simeq \tau_c v_c/c$ for an individual cluster, where the optical depth through the cluster is typically of order $\tau_c \simeq 0.1 - 0.01$ and its peculiar velocity is $v_p \simeq 500 - 1000$ km s$^{-1}$. Again there is hope that the signal can be removed by identifying bright clusters and perhaps even by first detecting the thermal effect. In any case the background fluctuations due to this effect are likely to be small [108].

### 5. Gravitational Lensing

Gravitational lensing processes the primary anisotropies by redistributing power in $\ell$. The magnitude and sense of the effect is somewhat dependent on the model for structure formation, including the assumptions for non-linear clustering. This has led to some seemingly inconsistent results in the literature (e.g. [109,110,111,112,113,114,115,116,117, 118,119,120]). Recently [121] has shown that for CDM, and indeed most realistic scenarios of structure formation, the effect is small on arcminute scales and above. Nonetheless, it is comparable to some of the minor corrections considered here. Lensing smooths out the primary peaks on the order of 5% at the third peak and becomes increasingly important toward smaller scales. Note, however, that since lensing conserves power, broad band measurements of anisotropies would not be able to measure this smoothing, unlike other effects considered here.

### 6. Other Scattering Effects

The standard Boltzmann equation includes only Thomson scattering as a source term. In principle there could be other scattering effects. If spectral distortions are present, due perhaps to some energy injection into the CMB, double Compton scattering and bremsstrahlung sources must be included and act to thermalize the distribution [122,123]. Rayleigh scattering off neutral hydrogen is unlikely to be important except for $\lambda \lesssim 0.05$ cm [9,124]). Similarly there could be effects caused by molecules. It has been pointed out that LiH can lead to rescattering damping (analogous to reionization damping) of anisotropies in the Rayleigh-Jeans region. Thus this effect would have a distinguishable spectral signature [125]. Due to uncertainties in the primordial Li abundance and moreover the formation and survival of LiH, its significance is unclear. As the authors note, the argument is best reversed: the CMB may be a sensitive probe of high redshift molecule formation.

Scattering and absorption by dust will also be an effect at the angular scale of galaxies and clusters (perhaps up to a few arc minutes), but will generally have a distinctive effect on the spectrum [126,103]. So the effects of extra-galactic dust are unlikely to be important on the scales relevant for the primary anisotropies, although they may be the major source at arcsecond scales. Of course there are also potentially great complications caused by confusing foreground sources in the Galaxy.

## V. PHYSICAL APPROXIMATIONS AND CALCULATIONAL TECHNIQUES

### A. Gauge and Initial Conditions

Our calculations have been carried out using two entirely independent Boltzmann codes. One of us (NS) has developed a code which was written using variables defined in the gauge independent formalism [127,128,129] in the total matter rest frame representation. Another of us (MW) more recently wrote a code based on the synchronous gauge [130,131,132]. Obviously, all observable quantities, in particular the $C_\ell$'s are gauge independent and thus only superhorizon scale perturbations allow gauge ambiguity. However, there has been some confusion in the literature about unphysical gauge modes (see [133,38,39] for discussions), so it is not entirely obvious that calculations carried out in two different gauges will agree perfectly, due to stimulation of such modes by numerical errors.

Such concern is perhaps enhanced by the fact that apparently different equations are evolved in each gauge (see e.g. [17]) and involve different numerical techniques to insure stability. In particular, the initial conditions for density perturbations take on a simpler form in some gauges than others, i.e. those defined with constant time hypersurfaces that follow the total matter or CDM. Furthermore, they must be established by a detailed balance of component fluctuations that stimulates the growing mode only, so that any gauge modes that exist are eliminated. This is especially important for isocurvature conditions where cancellation between component densities exist in the initial conditions. In practice, this is achieved by employing analytic solutions for the relevant gauge at early times when the fluctuations are outside the horizon.

A first important test of the robustness of anisotropy calculations is to demonstrate that two independent calculations can reproduce the same results for exactly the same input parameters. We found that when we tried to ensure that the only difference between the two calculations was the choice of gauge and numerical techniques, we could obtain results consistent to better than 0.5% in power for the first three peaks, better than 1% in power (0.5% in temperature) for $\ell \lesssim 2000$, rising to a few percent in the damping region. This level of agreement was found to persist as improvements in our initial assumptions were incorporated. A test of conformal Newtonian versus synchronous gauge techniques can be found in [17].

### B. Tight Coupling

Before the recombination process lowers the free electron density, the Compton scattering rate is sufficiently rapid that the photons and baryons are tightly coupled. This means that anisotropies in the photon distribution cannot be generated and the infinite hierarchy of $\ell$-modes may be replaced by a photon-baryon fluid approximation (the modes with $\ell \geq 2$ are exponentially damped). Providing one switches from tight coupling to explicit evolution early enough ($z \simeq 7{,}000$, or better, well before the photon diffusion length overtakes the wavelength of the perturbation), and keeps higher orders in the expansion in the Compton scattering time (e.g. [132]), one obtains identical $C_\ell$'s with a significant improvement in speed. By sacrificing some accuracy, the semi-analytic approach of [4] exploits this fact by keeping a second order tight coupling approximation all the way through last scattering.

Note that even ignoring recombination, as is appropriate for early reionization scenarios, the tight coupling limit eventually breaks down due to the drop in the free electron density from the expansion. Thus, tight coupling approximations should not be used to calculate reionized scenarios.

### C. Free Streaming

After last scattering, photons free stream to the observer on null geodesics, projecting power in the monopole and dipole at last scattering onto anisotropies today. It would thus seem unnecessary to solve numerically the whole hierarchy in $\ell$ until the present. This picture is complicated by the fact that photons still suffer non-trivial effects of time dependent gravitational redshift [41], often referred to as the integrated Sachs-Wolfe (ISW) effect. In Fig. 17a, we show the simplest approximation possible. Here we neglect the ISW contributions and furthermore consider all contributions from the rms fluctuation to project as a monopole (pure inhomogeneity) at last scattering. This results in two types of severe errors. The ISW effect contributes strongly to anisotropies near the first peak thus boosting the net power in anisotropies. Furthermore, the presence of a dipole due to the bulk motion of the photon-baryon fluid at last scattering alters the projection due to the angular dependence on the sky. Unless this is tracked properly, the phase relation of the peaks is altered [4]. Notice that this approximation, which neglects the relative phases of the monopole and dipole, can also erroneously lead to a small decrement at $\ell \simeq 20 - 40$.

Therefore, a better approximation is to keep angular information in the $\ell$ modes (see [8], eq. 4.5). If the ISW term were kept, this in principle leads to an exact solution of the Boltzmann equation (see e.g. [4] where the monopole, dipole and ISW term are kept). One can avoid the difficulty in keeping the ISW term in an $\Omega_0 = 1$ universe by noting that it vanishes after full matter domination. Thus for sufficiently late times, the free streaming approximation neglecting ISW should again be exact. The penalty in this case is that more $\ell$-modes have been generated and all of their relative phases must be kept in the free streaming approximation. In Fig. 17b, we show the effect of switching to this approximation at different redshifts. Notice that for $z = 800$ the retention of the phase relations eliminates the decrement of Fig. 17a before the first peak, but is in serious error due to the lack of the ISW term. It is not until $z \lesssim 100$ that the universe is sufficiently matter dominated for this approximation to hold at the few percent level. It should be mentioned that this epoch depends on $\Omega_0 h^2$. We can ignore the ISW effect much earlier if we choose models with higher $h$.

Earlier work (e.g. [18,10,13,31] concentrated on calculations of $C(\theta)$, which tended to involve approximations which are only valid at small angles. Obviously such approximate methods fail to calculate $C(\theta)$ accurately when there is a reasonable amount of large scale power, which essentially means all models now discussed [8]. The small-angle formulae have been extended to include the Sachs-Wolfe effect [8]. However, this still is only an approximation, neglecting angular information and ISW contributions and will be in error by the same amounts as shown in Fig. 17a.

### D. Smoothing and $k$ Range

The free streaming of photons after last scattering causes rapid oscillations in $\ell$ of the anisotropy for any given $k$-mode. Physically this arises since the projection of plane waves onto the sphere at last scattering causes an aliasing of power to larger angles. Of course, any realistic model involves a spectrum of uncorrelated $k$-modes leading to a smoothly varying total anisotropy. In practice however, a finite number of $k$-modes are evolved, which causes noise in the theoretical predictions, i.e. spurious oscillations in $\ell$. Despite this it is not necessary to compute as many $k$ modes as $\ell$ modes (which would lead to smooth power spectra) in order to obtain accurate results for the $C_\ell$'s. This is because the error in the spectrum which comes from finite sampling in $k$-space simply induces a 'shot noise' on top of the underlying smooth spectrum which can be accurately removed by low pass filtering the resulting $C_\ell$'s. We show an example of this filtering process in Fig. 18. The top left panel shows the angular power spectrum obtained from evolving 500 $k$-modes, with no smoothing. At the right is the FFT of this spectrum (squared). Note the peaks in power at $\nu = 600, 1200, ...$ which are the 'noise' we wish to remove. The noise is separated in frequency by a range near $\nu = 100$ where little power exists. This separation corresponds to the gap between the $\ell$ to $\ell$ variation of the noise and the size of the smallest physical feature in the power spectrum. The bottom right panel shows the low-pass filtered FFT. Bottom left is the angular power spectrum obtained by FFTing the bottom right panel. Note that the noise has been efficiently removed. Comparison with models in which more $k$-modes are run to remove the noise shows that it is possible to recover the underlying spectrum to much better than 0.5%. This plot is shown for pedagogical purposes only, and is not the best smoothing. In practice we sample the spectrum with different weightings in $\ell$ in order to recover the different parts of the spectrum.

There are probably optimal ways of choosing the spacing of the $k$-modes in order to recover the most accurate anisotropy spectrum with the minimum redundancy. We have not investigated this fully, although somewhere between $\log k$ and $k$ spacing seems clearly to be best. Similarly there are better or worse ways of choosing the spacings to use in the FFT (going from $\log \ell$ to $\ell^{1/2}$ or $\ell$ works well), and perhaps more sophisticated methods of smoothing which adequately preserve the structure in the $C_\ell$'s for the smallest number of $k$ calculations. However, these are not a serious consideration for the current level of achievable accuracy.

Thus as long as sufficient $k$ modes (usually $\simeq 500$) are kept to insure that the noise varies on scales smaller than real minimum-width features of the power spectrum, this results in accurate reconstruction of the underlying power spectrum, as can be checked by increasing the number of $k$-modes evolved.

Of course, these $k$-modes are chosen to span a range corresponding to the desired angular limit of the calculation. The projection of spatial scales on the last scattering surface (or more generally at the epoch of anisotropy generation) onto angles on the sky today, leads to a relation between the maximum $k$-mode of the calculation and the maximum $\ell$-mode at which accurate values for the anisotropy are required. We find it is necessary to keep $k$ up to $2\ell/\eta_0$ to get sufficient accuracy, where $\eta_0$ is the present conformal time which is $6000 h^{-1}$Mpc for $\Omega_0 = 1$. The error made by truncating the calculation at high $k$ is shown in Fig. 19. Here we have calculated the power spectrum using more and more $k$ modes, comparing the result to a calculation which has a maximum $k$ of $0.8 h\,\text{Mpc}^{-1}$.

In summary, for full accuracy in the damping region, a large range of $k$-modes is required. However, these modes are numerically very expensive to compute and such investment in computing time may not be worthwhile, since extra physical effects make the high-$\ell$ anisotropies uncertain regardless.

### E. Tilt

Since all $k$-modes in linear theory evolve independently, the evolved $k$-space power spectrum can be factored into the initial power spectrum times a transfer function that accounts for the evolution. The full transfer function for the photons must of course be two dimensional to account for the angular power in $\ell$ (see e.g. [5]). However, since most of the effect is a projection of a particular linear scale onto an angular scale, the transfer function is nearly a one to one mapping between $\ell$'s and $k$'s for fluctuations originating at a single epoch, e.g. the last scattering surface. It is thus possible to estimate the effect of varying $n$ in an $\Omega_0 = 1$ universe by applying a transfer function in $\ell$ alone. In Fig. 20, we approximate tilt by multiplying the $C_\ell(n=1)$ by the Sachs-Wolfe $\Gamma$ function formula, which approximately takes into account the width in the $\ell$ to $k$ mapping [134]. This approximation works at the $\lesssim 5\%$ level for $n \gtrsim 0.9$. Multiplying the $C_\ell$ by $\ell^{n-1}$ is only slightly worse. For cases where the ISW term contributes significantly and anisotropy generation is spread out significantly in time, these approximations break down entirely. An alternative approach has been discussed in [6] which works better for $\Lambda$ models.

Note also that the effective slope $n$ is not necessarily constant over the relevant range of $\ell$: the $(\log(k/k_0))^3$ correction which occurs in some inflationary models deviates from an $n = 0.945$ (best fit) spectrum at the 2% level over 3 orders of magnitude in $k$ (or $\ell$). So even this 'minimal' departure from $n = 1$ may give an observable effect. There is also the possibility of detecting some more non-trivial deviation from power-law initial conditions by measuring accurate $C_\ell$'s.

### VI. CONCLUSIONS

In this paper, we have only considered one class of models, namely the standard Cold Dark Matter model. Nevertheless, we expect that the sizes of the effects will be fairly general so that the spirit of our conclusions will apply to other models. However, we have completely neglected possibilities such as the inclusion of a cosmological constant, open models, isocurvature fluctuations or cosmological defects. It is entirely possible that in such models some of the effects we have talked about could be either more or less important. However, one can carry out such tests for any other specific class of theory, and we hope that our results will be a guide to the kinds of effects which need to be considered if theorists are to perform calculations to the level of accuracy that the new generation of experiments are demanding.

The potential power of future experiments is truly awe-inspiring. It will be possible in principle to measure all $C_\ell$'s up to $\ell \sim 500$ to the cosmic variance limit (i.e. accuracy $\sim 1/\sqrt{\ell}$). In practice it may be possible to achieve close to this ideal, depending how restrictive foreground contamination proves to be. The amount of information coded in the $C_\ell$'s is enormous. In order to obtain constraints on the standard cosmological parameters $\Omega_0, \Omega_B, h, \Lambda, n, T/S$ etc., it will be necessary to consider in detail all of the effects discussed here. Moreover, there will be the possibility of extracting information about more subtle physical effects, e.g. the number of relativistic species, or the curvature of the primordial power spectrum. At the highest $\ell$'s (where extracting foregrounds may be the most difficult), there will be multiple, complicated effects. It will be a great challenge to disentangle the various physical processes out towards the damping region, to get at the reionization history, the non-linear evolution etc. We hope that our work here has been a step towards obtaining anisotropy power spectra that are accurate enough to enable us to be up to the task when the next generation of data are available.

## ACKNOWLEDGMENTS

We would like to thank Dick Bond, Ted Bunn, Scott Dodelson, Kris Górski, Marc Kamionkowski, Jim Peebles, Uroš Seljak, Joe Silk, David Spergel, Paul Steinhardt and Radek Stompor for helpful conversations. We are grateful for the hospitality of the ITP where the final parts of this research were completed. This research was supported in part by the National Science Foundation under Grants No. AST91-20005, AST92-04639 and PHY89-04035. NS acknowledges financial support from a JSPS postdoctoral fellowship for research abroad. WH acknowledges the support of an NSF fellowship.

## REFERENCES

[1] M. White, D. Scott, and J. Silk, Ann. Rev. Astron. Astroph. **32**, 319 (1994).

[2] J. R. Bond, R. Crittenden, R. L. Davis, G. Efstathiou, and P. J. Steinhardt, Phys. Rev. Lett. **72**, 13 (1994).

[3] U. Seljak, Astrophys. J. **435**, L87 (1994).

[4] W. Hu and N. Sugiyama, Astrophys. J. in press (1995).

[5] W. Hu and N. Sugiyama, Phys. Rev. D. **51**, 2599 (1995).

[6] N. Sugiyama preprint astro-ph 9412025 (1994).

[7] R. Stompor, Astron. Astrophys. **287**, 693 (1994).

[8] J. R. Bond and G. Efstathiou, Mon. Not. R. Astron. Soc **226**, 655 (1987).

[9] P. J. E. Peebles and J. T. Yu, Astrophys. J. **162**, 815 (1970).

[10] M. L. Wilson and J. Silk, Astrophys. J. **243**, 14 (1981).

[11] P.J. E. Peebles, Astrophys. J. **248**, 885 (1981).

[12] J. R. Bond and A. Szalay, Astrophys. J. **274**, 443 (1983).

[13] J. R. Bond and G. Efstathiou, Astrophys. J. **285**, L45 (1984).

[14] W. Hu, D. Scott, and J. Silk, Phys. Rev. D. **49**, 648 (1994).

[15] S. Dodelson, and J. M. Jubas, Astrophys. J. **439**, 503 (1995).

[16] J. Bernstein and S. Dodelson, Phys. Rev. D. **41**, 354 (1990).

[17] C.-P. Ma and E. Bertschinger,, Astrophys. J. in press (1994).

[18] A. G. Doroshkevich, Ya. B. Zel'dovich, and R. A. Sunyaev, Sov. Astron. **22**, 523 (1978).

[19] S. A. Bonometto, A. Caldara, and F. Lucchin, Astron. Astrophys. **126**, 377 (1983).

[20] S. A. Bonometto, F. Lucchin, and R. Valdarnini, Astron. Astrophys. **140**, L27 (1984).

[21] B. J. T. Jones and R. F. G. Wyse, Mon. Not. R. Astron. Soc **205**, 983 (1983).

[22] A. A. Starobinskiĭ, Sov. Astron. Lett. **14**, 166 (1988).

[23] A. G. Doroshkevich, Sov. Astron. Lett. **14**, 125 (1988).

[24] F. Atrio-Barandela, A. G. Doroshkevich, and A. A. Klypin , Astrophys. J. **378**, 1 (1991).

[25] H. E. Jørgensen, E. V. Kotok, P. D. Naselsky, and I. D. Novikov In: *Present, and Future of the Cosmic Microwave Background*, 1994 edited by J. L. Sanz, E. Martínez-González, and L. Cayón, Springer-Verlag, Berlin, p. 146.

[26] H. E. Jørgensen, E. V. Kotok, P. D. Naselsky, and I. D. Novikov, Astron. Astrophys. in press (1995).

[27] N. Kaiser, Astrophys. J. **282**, 374 (1984).

[28] E. T. Vishniac, Astrophys. J. **322**, 597 (1987).

[29] G. Efstathiou In *Large-Scale Motions in the Universe. A Vatican Study Week*, 1988 edited by V. C. Rubin

and G. V. Coyne, Princeton University Press, Princeton N.J., p. 299.

[30] F. Atrio-Barandela and A. G. Doroshkevich, Astrophys. J. **420**, 26 (1994).

[31] N. Vittorio and J. Silk, Astrophys. J. **285**, L39 (1984).

[32] N. Vittorio and J. Silk, Astrophys. J. **385**, L9 (1992).

[33] J. A. Holtzman, Astrophys. J. Suppl. **71**, 1 (1989).

[34] N. Sugiyama and N. Gouda, Prog. Theor. Phys. **88**, 803 (1992).

[35] S. Dodelson and J. M. Jubas, Phys. Rev. Lett. **70**, 2224 (1993).

[36] R. Crittenden, J. R. Bond, R. L. Davis, G. Efstathiou, P. J. Steinhardt, Phys. Rev. Lett. **71**, 324 (1993a).

[37] R. Crittenden, R. L. Davis, P. J. Steinhardt, Astrophys. J. **417**, L13 (1993b).

[38] J. R. Bond In *The Early Universe*, 1988 edited by W. G. Unruh and G. W. Semenoff, Dordrecht: Reidel, p. 283.

[39] G. Efstathiou In *Physics of the Early Universe: Proceedings of the 36th Scottish Universities Summer School in Physics*, 1990 edited by J. A. Peacock, A. E. Heavens, and A. T. Davies, Adam Hilger, New York, p. 361

[40] J. C. Mather *et al.*, Astrophys. J. **420**, 439 (1994).

[41] R. K. Sachs and A.M Wolfe, Astrophys. J. **147**, 73 (1967).

[42] M. White, G. Gelmini, and J. Silk, Phys. Rev. D **51**, 2669 (1995).

[43] M. Kamionkowski and D. N. Spergel private communication (1994).

[44] U. Seljak and E. Bertschinger In: *Present and Future of the Cosmic Microwave Background*, 1994 edited by J. L. Sanz, E. Martínez-González, and L. Cayón, Springer-Verlag, Berlin, p. 164.

[45] S. Dodelson *et al.* in preparation (1995).

[46] D. E. Osterbrock *Astrophysics of Gaseous Nebulae and Active Galactic Nuclei*, 1989 University Science Books; Mill Valley, CA.

[47] I. D. Novikov, Ya. B. Zel'dovich, Ann. Rev. Astron. Astrophys. **5**, 627 (1967).

[48] P. J. E. Peebles, Astrophys. J. **153**, 1 (1968).

[49] Ya. B. Zel'dovich, V. G. Kurt, and R. A. Sunyaev, Zh. Eksp. Teor. Fiz. **55**, 278, English translation: *Sov. Phys. – JETP*, 28:146 (1969) (1968).

[50] B. J. T Jones and R. F. G. Wyse, Astron. Astrophys. **149**, 144 (1985).

[51] S. I. Grachev and V. K. Dubrovich, Astrophys. **34**, 124 (1991).

[52] T. Matsuda, H. Sato, and H. Takeda, Prog. Theor. Phys. **42**, 219 (1969).

[53] T. Matsuda, H. Sato, and H. Takeda, Prog. Theor. Phys. **46**, 416 (1971).

[54] Y. E. Lyubarsky and R. A. Sunyaev, Astron. Astrophys. **123**, 171 (1983).

[55] P. J. E. Peebles private communication (1995).

[56] J. R. Bond private communication (1995).

[57] R. Weymann, Astrophys. J. **145**, 560 (1966).

[58] H. Sato, T. Matsuda, and H. Takeda, Prog. Theor. Phys. Suppl. **49**, 11 (1971).

[59] J. H. Krolik, Astrophys. J. **338**, 594 (1989).

[60] J. H. Krolik, Astrophys. J. **353**, 21 (1990).

[61] G. B. Rybicki and I. P. Dell'Antonio, Astrophys. J. **427**, 603 (1994).

[62] V. V. Burdyuzha and A. N. Chekmezov, Astron. Rep. **38**, 297 (1994).

[63] S. Sasaki and F. Takahara, Pub. Astron. Soc. of Japan **45**, 655 (1993).

[64] C. W. Allen *Astrophysical Quantities*, 1973 Third Edition, The Athlone Press, University of London.

[65] D. G. Hummer and M. J. Seaton, Mon. Not. R. Astron. Soc **125**, 437 (1963).

[66] D. G. Hummer, Mon. Not. R. Astron. Soc **268**, 109 (1994).

[67] D. Péquignot, P. Petitjean, and C. Boisson, Astron. Astrophys. **251**, 680 (1991).

[68] L. Spitzer *Physical Processes in the Interstellar Medium*, 1978 Wiley, New York.

[69] S. P. Goldman, Phys. Rev. A **40**, 1185 (1989).

[70] A. Kosowsky astro-ph 9501045 (1995).

[71] M. J. Rees, Astrophys. J. **153**, L1 (1968).

[72] N. Kaiser, Mon. Not. R. Astron. Soc **202**, 1169 (1983).

[73] A. A. Starobinskiǐ, JETP Lett. **30**, 682 (1979).

[74] R. Fabbri and M. Pollock, Ap. Phys. Lett. **B125**, 445 (1983).

[75] L. F. Abbott, M.B. Wise, Nucl. Phys. **B244**, 541 (1984).

[76] A.A. Starobinskiǐ, Sov. Astron. Lett. **11**, 113 (1985).

[77] B. Allen and S. Koranda, Phys. Rev. D **50**, 3713 (1994).

[78] A.G. Polnarev, Sov. Astron. **29**, 607 (1985).

[79] D. Coulson, R.G. Crittenden, and N. Turok, Phys. Rev. Lett. **73**, 2390 (1994).

[80] R. G. Crittenden, D. Coulson, and N. Turok astro-ph 9411107 (1994).

[81] M. Tegmark, J. Silk, and A. Blanchard, Astrophys. J. **434**, 395 (1994).

[82] A. R. Liddle and D. H. Lyth, Mon. Not. R. Astron. Soc, in press (1995).

[83] R. A. Sunyaev, Sov. Astron. Lett. **3**, 268 (1977).

[84] N. Sugiyama, J. Silk, and N. Vittorio, Astrophys. J. **419**, L1 (1993).

[85] J. E. Gunn and B. A. Peterson, Astrophys. J.L **318**, L11 (1965).

[86] D. P. Schneider, M. Schmidt, and J. E. Gunn, Astron. J. **98**, 1951 (1989).

[87] J. K. Webb, X. Barcons, R. F. Carswell, and H. C. Parnell, Mon. Not. R. Astron. Soc. **255**, 319 (1992).

[88] J. P. Ostriker and E. T. Vishniac, Astrophys. J. **306**, L51 (1986).

[89] G. Efstathiou and J. R. Bond, Mon. Not. R. Astron. Soc **227**, 33p (1987).

[90] T. Chiba, N. Sugiyama, and Y. Suto, Astrophys. J. **429**, 427 (1994).

[91] W. Hu and M. White in preparation (1995).

[92] M. J. Rees and D. W. Sciama, Nature **517**, 611 (1968).

[93] R. Tuluie and P. Laguna *astro-ph*, 1995 9501059.

[94] E. Martinez-Gonzalez, J.L. Sanz, and J. Silk, Astrophys. J. **436**, 1 (1994).

[95] U. Seljak in preparation (1995).

[96] R. A. Sunyaev and Ya. B. Zel'dovich, Astrophys. Sp. Sci. **7**, 1 (1970).

[97] V. A. Korolev, R. A. Sunyaev, and L. A. Yakubtsev, Sov. Astron. Lett. **12**, 141 (1986).

[98] A. Cavaliere and S. Colafranceso, Astrophys. J. **415**, 50 (1988).

[99] S. Cole and N. Kaiser, Mon. Not. R. Astron. Soc **233**, 637 (1989).

[100] R. Schaeffer and J. Silk, Astrophys. J. **333**, 509 (1988).

[101] M. Markevitch, G. R. Blumenthal, W. Forman, C. Jones, and R. A. Sunyaev, Astrophys. J. **378**, L33 (1991).

[102] M. Markevitch, G. R. Blumenthal, W. Forman, W., C. Jones, and R. A. Sunyaev, Astrophys. J. **395**, 326 (1992).

[103] J. R. Bond and S. Myers *Trends in Astroparticle Physics*, 1991 edited by D. Cline and R. Peccei, World Scientific, Singapore, p. 262.

[104] N. Makino and Y. Suto, Astrophys. J. **405**, 1 (1993).

[105] J. G. Bartlett and J. Silk, Astrophys. J. **423**, 12 (1994).

[106] S. Colafrancesco, P. Mazzotta, Y. Rephaeli, and N. Vittorio, Astrophys. J. **433**, 454 (1994).

[107] M. T. Ceballos and X. Barcons astro-ph 9405052 (1994).

[108] F. M. Persi, D. N. Spergel, R. Cen, and J. P. Ostriker, Astrophys. J. **442**, 1 (1995).

[109] A. Blanchard and J. Schneider, Astron. Astrophys. **184**, 1 (1987).

[110] A. Kashlinsky, Astrophys. J. **331**, L1 (1988).

[111] S. Cole and G. Efstathiou, Mon. Not. R. Astron. Soc **239**, 195 (1989).

[112] M. Sasaki, Mon. Not. R. Astron. Soc **240**, 415 (1989).

[113] K. Tomita and K. Watanabe, Prog. Theor. Phys. **82**, 563 (1989).

[114] V. E. Linder, Mon. Not. R. Astron. Soc **243**, 353 (1990a).

[115] V. E. Linder, Mon. Not. R. Astron. Soc **243**, 362 (1990b).

[116] L. Cayón, E. Martínez-González, and J. L. Sanz, Astrophys. J. **403**, 471 (1993a).

[117] L. Cayón, E. Martínez-González, and J. L. Sanz,, Astrophys. J. **413**, 10 (1993b).

[118] L. Cayón, E. Martínez-González, and J. L. Sanz, Astron. Astrophys. **284**, 719 (1994).

[119] B. A. C. C. Bassett, P. K. S. Dunsby, and G. F. R. Ellis *Phys. Rev. D.*, 1994 submitted.

[120] T. Fukushige, J. Makino, and T. Ebisuzaki, Astrophys. J. in press (1994).

[121] U. Seljak, Astrophys. J. in press (1995).

[122] C. Burigana, L. Danese, and G. De Zotti, Astron. Astrophys **246**, 59 (1991).

[123] W. Hu and J. Silk, Phys. Rev. D **48**, 485 (1993).

[124] F. Takahara and S. Sasaki, Prog. Theor. Phys. **86**, 1021 (1991).

[125] R. Maoli, F. Melchiorri, and D. Tosti, Astrophys. J. **425**, 372 (1994).

[126] J. R. Bond, B. J. Carr, and C. J. Hogan, Astrophys. J. **367**, 420 (1991).

[127] J.M. Bardeen, Phys. Rev. D **22**, 1882 (1980).

[128] H. Kodama and M. Sasaki, Prog. Theor. Phys. Suppl. **78**, 1 (1984).

[129] V. F. Mukhanov, H. A. Feldman, and R. H. Brandenberger, Phys Rep **215**, 203 (1992).

[130] E. M. Lifshitz, JETP Lett. **16**, 587 (1946).

[131] L. D. Landau and E. M. Lifshitz *The Classical Theory of Fields*, 1975 4th Edition, Pergamon Press, Oxford.

[132] P. J. E. Peebles *The Large-Scale Structure of the Universe*, 1980 Princeton University Press, Princeton, N.J. §81.

[133] W. Press and E. Vishniac, Astrophys. J. **239**, 1 (1980).

[134] M. White, L. M. Krauss, and J. Silk, Astrophys. J. **418**, 535 (1992).

# FIGURE CAPTIONS

FIG. 1 The angular power spectrum of CMB temperature anisotropies, i.e. $\ell(\ell+1)C_\ell/6C_2$ versus $\ell$, for our fiducial model (multipole $\ell \sim \theta^{-1}$). This includes all the effects that are discussed in this paper, and hence we expect this plot to be accurate to $\lesssim 1\%$ until the damping region at $\ell \gtrsim 1500$. Also shown is the polarization power spectrum, normalized to the temperature quadrupole.

FIG. 2 The relative error in $C_\ell$ for different choices of the present day photon temperature, $T_{\gamma 0}$. Plotted is the ratio of $C_\ell$ for $T_{\gamma 0} = 3\,\mathrm{K}$ (dotted), $2.7\,\mathrm{K}$ (solid), and $2.73\,\mathrm{K}$ (dashed), relative to $T_{\gamma 0} = 2.726\,\mathrm{K}$.

FIG. 3 (a) The angular power spectrum of CMB anisotropies with varying number of neutrino species: $N_\nu = 2, 3, 4$. (b) The ratio of $C_\ell$ for $N_\nu = 2, 4$ relative to $N_\nu = 3$.

FIG. 4 The effect on the $C_\ell$'s of a small increase in $\Omega_B h^2$ with $h$ fixed at 0.5.

FIG. 5 The effect on the $C_\ell$'s of a small increase in $h$ with $\Omega_B h^2$ fixed at 0.0125.

FIG. 6 The ratio of $C_\ell$ for models with primordial helium $Y_P = 0.20$ and $Y_P = 0.26$, relative to our fiducial model with $Y_P = 0.23$.

FIG. 7 The effect of different approximations for the physics of recombination. The solid line shows the ratio of $C_\ell$ using accurate values for the recombination coefficients $\alpha(T)$, relative to a model using values which scale as $T^{-1/2}$. The dashed curve shows the extra effect on the $C_\ell$'s of adding the helium recombination.

FIG. 8 The recombination of the plasma for a model with the parameters of standard CDM (i.e. $\Omega = 1$, $\Omega_B = 0.05$, $h = 0.5$) and with $Y_P = 0.23$. The quantity plotted on the y-axis is $n_e/n_H$, which is unity when the hydrogen is totally ionized, and is higher by an amount $2n_{He}/n_H$ when the helium is doubly ionized too.

FIG. 9 The effect on the $C_\ell$'s of adopting the assumption of simple Saha recombination (dashed), compared with a more accurate recombination calculation.

FIG. 10 Hydrogen recombination. (a) Ionization history, $x_e(z)$, using the naive coefficients $\alpha \sim T^{1/2}$ (dotted) and more accurate fitting functions (all of which are indistinguishable from the dashed line). Also shown is the result of following the kinetic temperature of the matter explicitly (solid). The lowest dashed line has $z_{\mathrm{eq}} \to \infty$, i.e. ignoring the effect of the radiation on the background evolution. (b) Visibility functions ($g(z) = e^{-\tau} d\tau/dz$) for the same recombination approximations.

FIG. 11 The angular power spectrum calculated including the neutrino anisotropy explicitly (solid) and treating the neutrinos as a perfect fluid only, i.e. truncating the hierarchy after $\ell = 1$ (dashed).

FIG. 12 The spectrum of anisotropies for the background neutrinos themselves in a standard CDM model. The amplitude at small $\ell$ is the same as for the photons. Scales which enter the horizon during the radiation-dominated epoch have an additional integrated Sachs-Wolfe contribution.

FIG. 13 Polarization. The ratio of $C_\ell$ for a standard CDM model where polarization is explicitly followed, relative to a calculation where it is neglected. Notice that the polarization anisotropy has a non-negligible feed-back effect for the temperature anisotropies.

FIG. 14 Gravity wave spectrum. $\ell(\ell+1)C_\ell$ versus $\ell$ for a flat spectrum of tensor-generated anisotropies with the parameters of the standard CDM model. In general this spectrum is added incoherently to the scalar one (Fig. 1), with the relative amplitude determined by the specific inflationary model. Since the tensor spectrum drops rapidly for $\ell \gtrsim 100$ most of the effects described in this paper are not important for obtaining sufficiently accurate calculations.

FIG. 15 Minimal reionization. The ratio of $C_\ell$ for a model with reionization at $z = 4$ (as required by results of the Gunn-Peterson test) compared with our fiducial model. We assume full ionization of the Hydrogen and (conservatively) that the Helium remains totally neutral.

FIG. 16 The complexity of realistic reionization scenarios. Ratios of $C_\ell$ for two different reionization histories compared to our fiducial model. The scenarios assume standard recombination until a redshift $z_{\rm ri} = 30(48)$ below which the electron fraction is set to a constant $x_e = 1(0.5)$. These models both have the same optical depth, $\tau = 0.15$, back to redshift $z \sim 100$. Also shown is a crude approximation where the $C_\ell$ are reduced by $\exp(-2\tau_\ell)$ where $\tau_\ell = \tau(z(\theta = \pi/\ell))$. Here $\tau(z)$ is the optical depth from redshift 0 to $z$, we use the angle subtended by the horizon to map angles to redshifts, and $\theta = \pi/\ell$ to map multipoles to angles.

FIG. 17 Free Streaming approximations. (a) Streaming of the rms fluctuations from epochs $z = 50$, 100, 200, 300, 500 and 800, neglecting the ISW contributions. The solid line is the result of integrating to the present. This shows that it is not possible to stop the calculation at some redshift and take a 'snapshot' of the inhomogeneities on this surface. (b) Streaming of *all* multipoles from the same epochs, but still neglecting ISW. Even including all the moments of the photon anisotropies at some early redshift surface is not enough to accurately reproduce the present day anisotropies.

FIG. 18 Low pass filter smoothing of the $C_\ell$'s. Clockwise from top left: (1) noisy spectrum due to 'sparse' sampling in $k$; (2) Fourier transform power; (3) low pass filtered FFT; (4) smoothed spectrum obtained by FFT of (3). Note that this figure is illustrative only. More complicated sampling of the x-axis allows better reconstruction of the high $\ell$ peaks, as in Fig. 1.

FIG. 19 $k$-range. We show the ratio of $C_\ell$ difference to our fiducial model (with $k\eta_0 = 5000$, for a set of calculations with the highest $k$ kept being 800 to 4500, reading left to right). Plotted (dashed) on top is the actual $C_\ell$'s to show the position of the damping and other features.

FIG. 20 Tilt. The ratio of the $C_\ell$'s using the tilt approximation described in the text, to the exact calculation, for $n = 0.85, 0.9, 0.95$. This approximation is adequate for comparison with today's experiments, but not accurate at the 1% level.

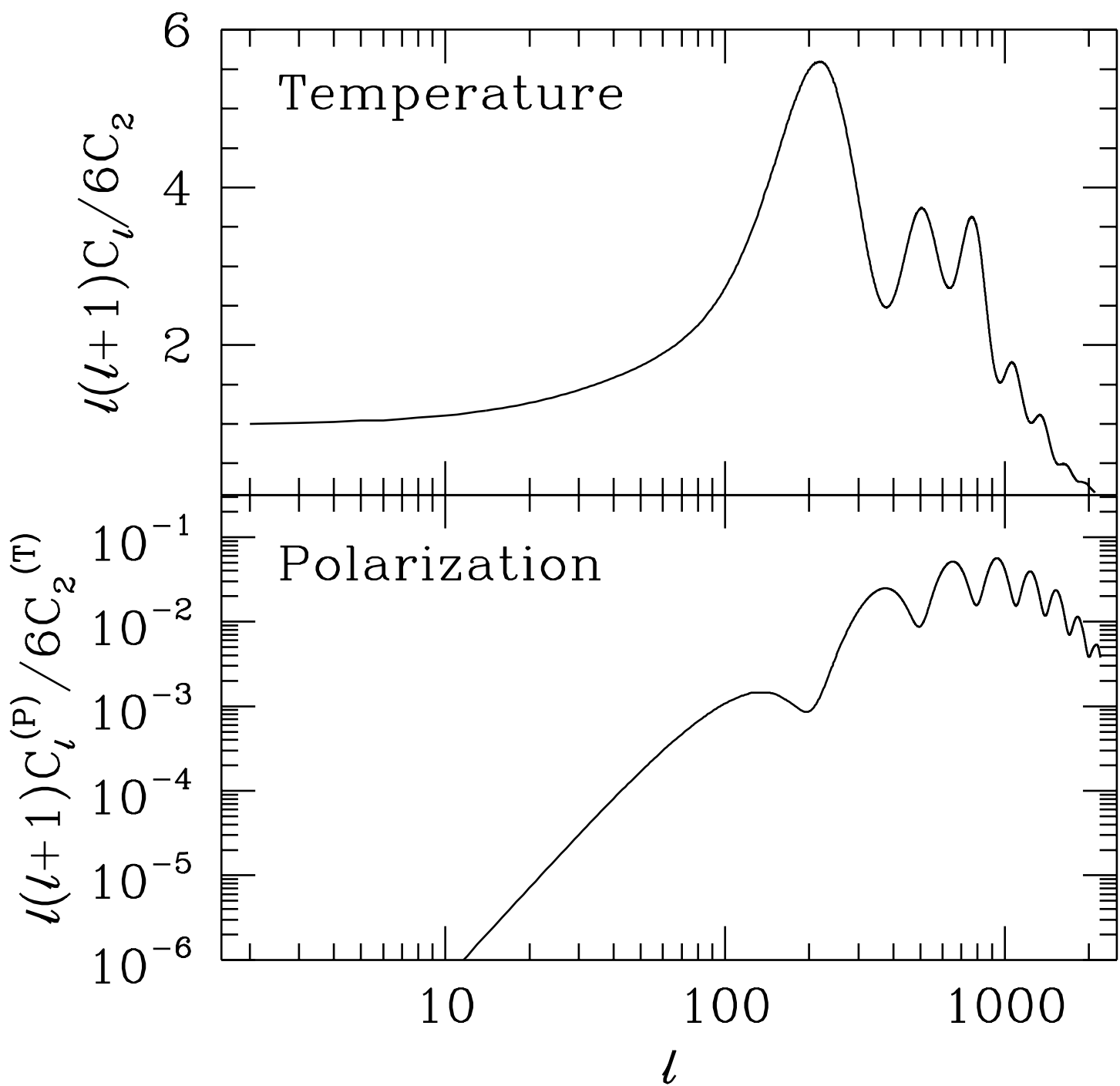

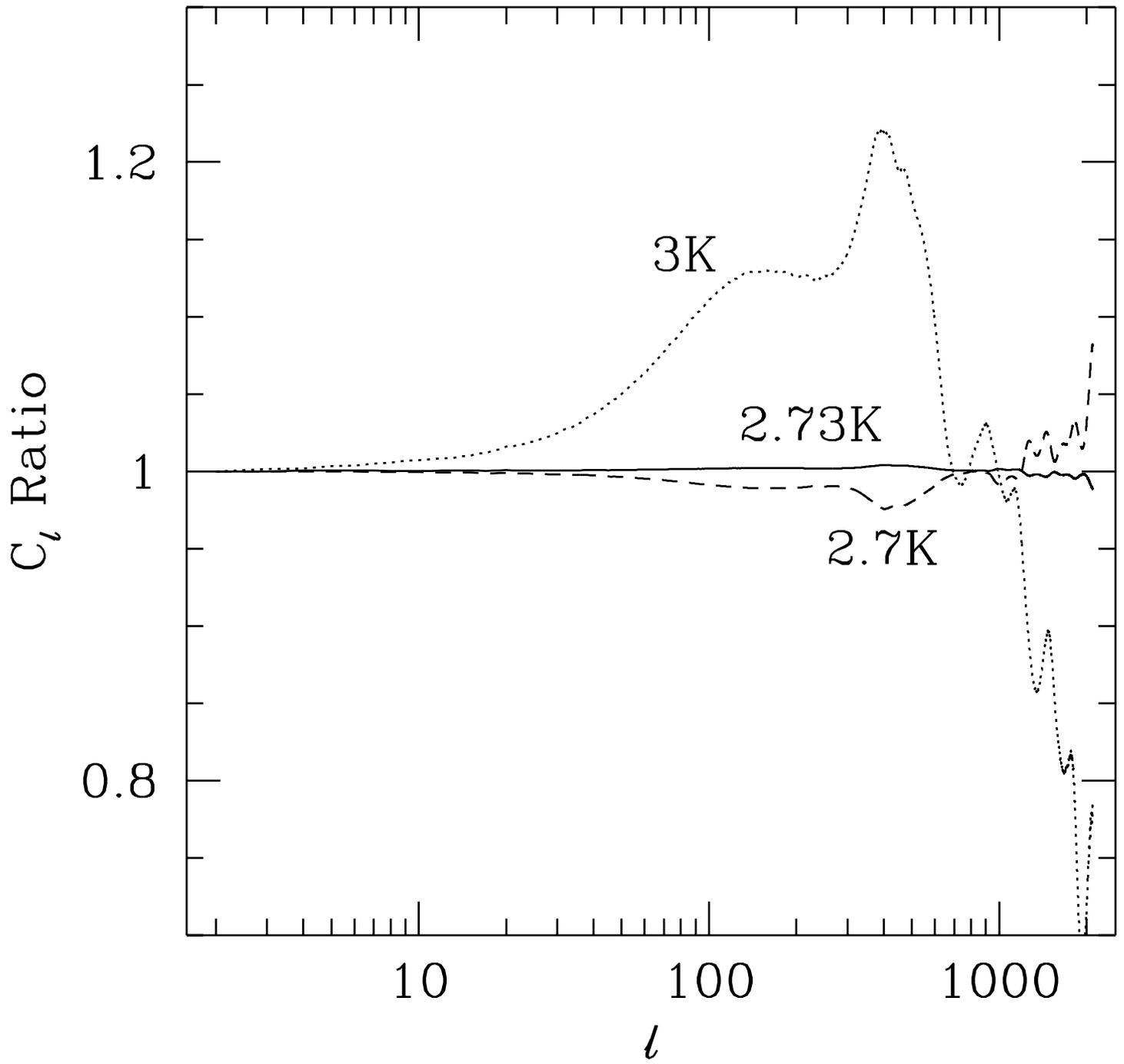

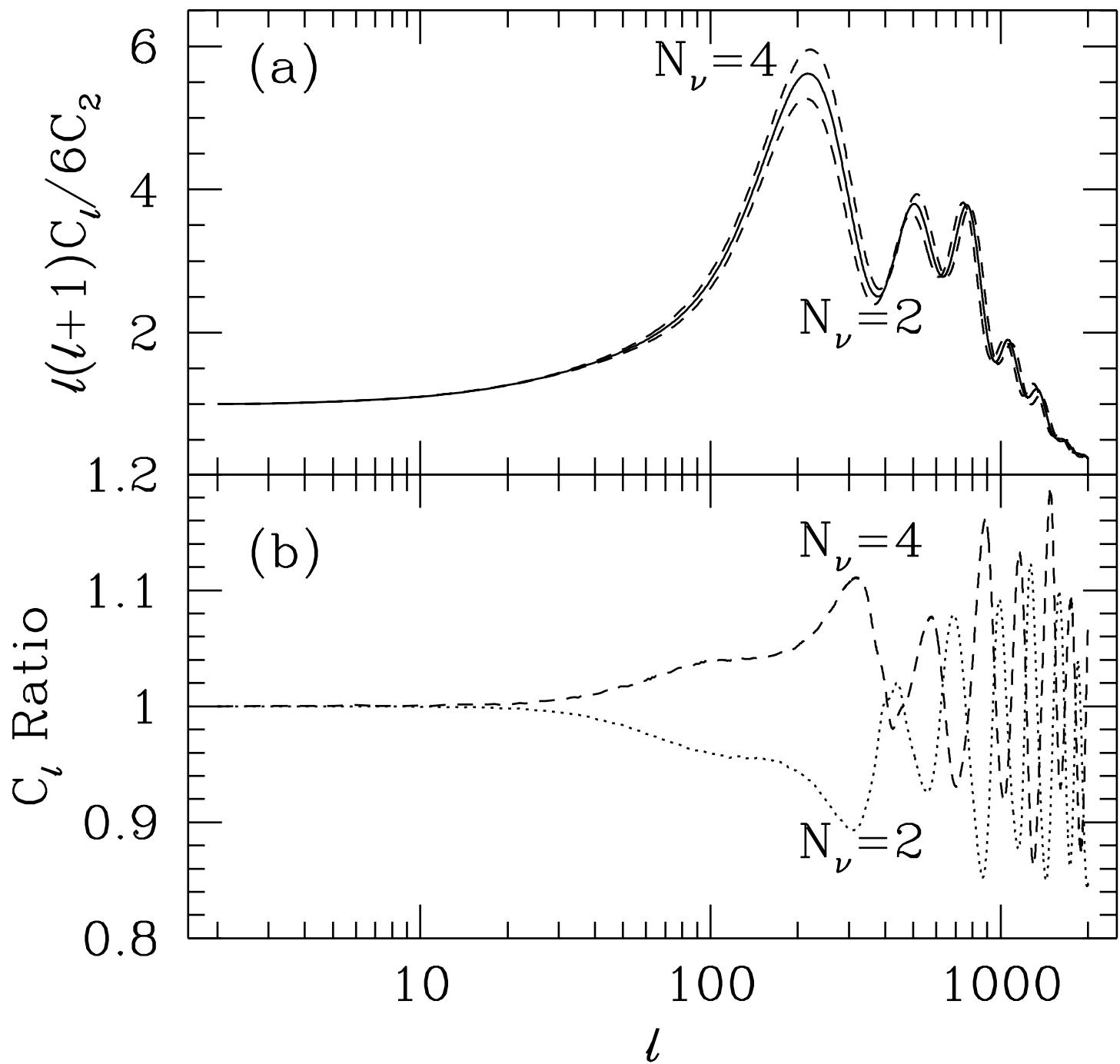

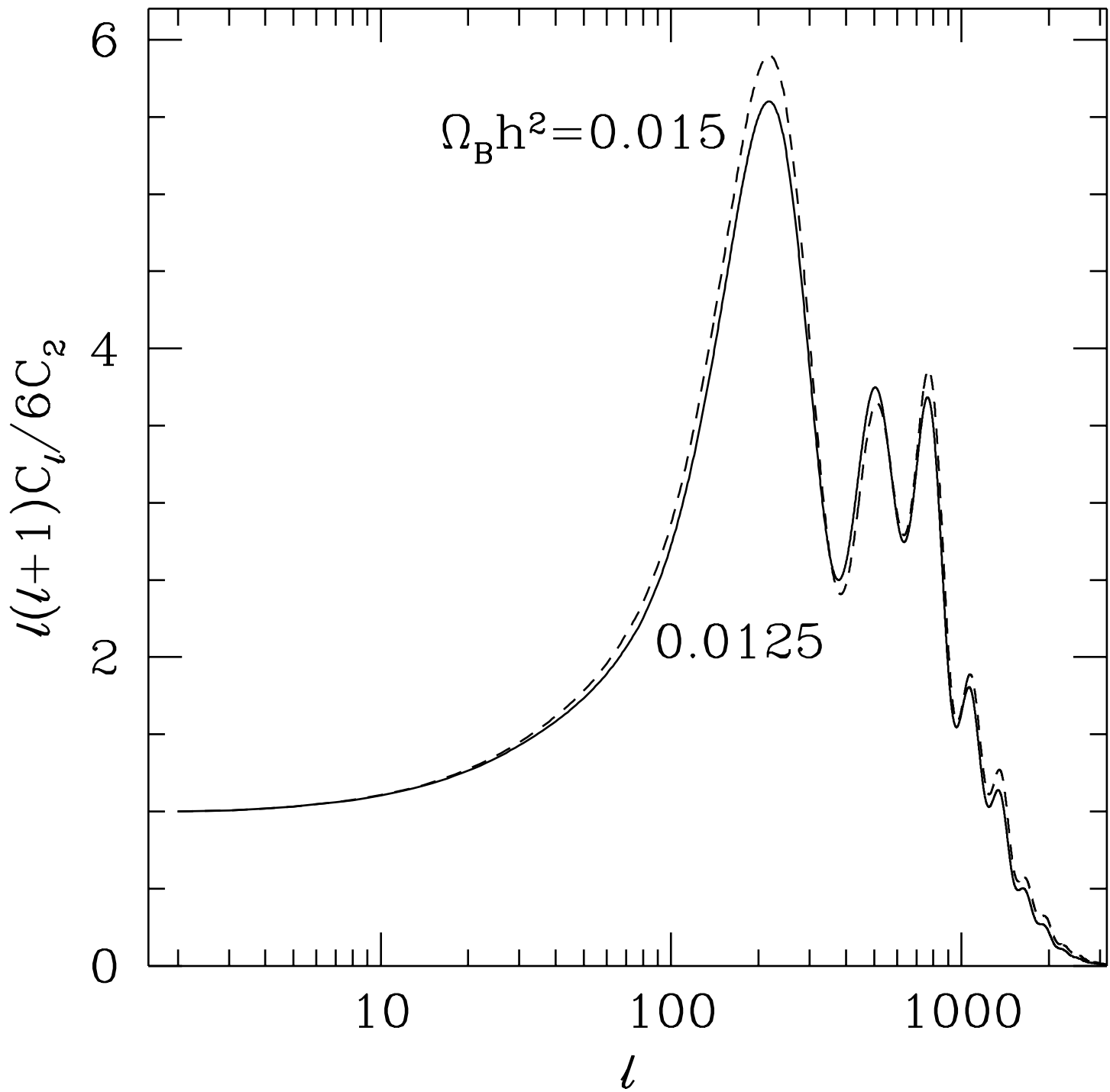

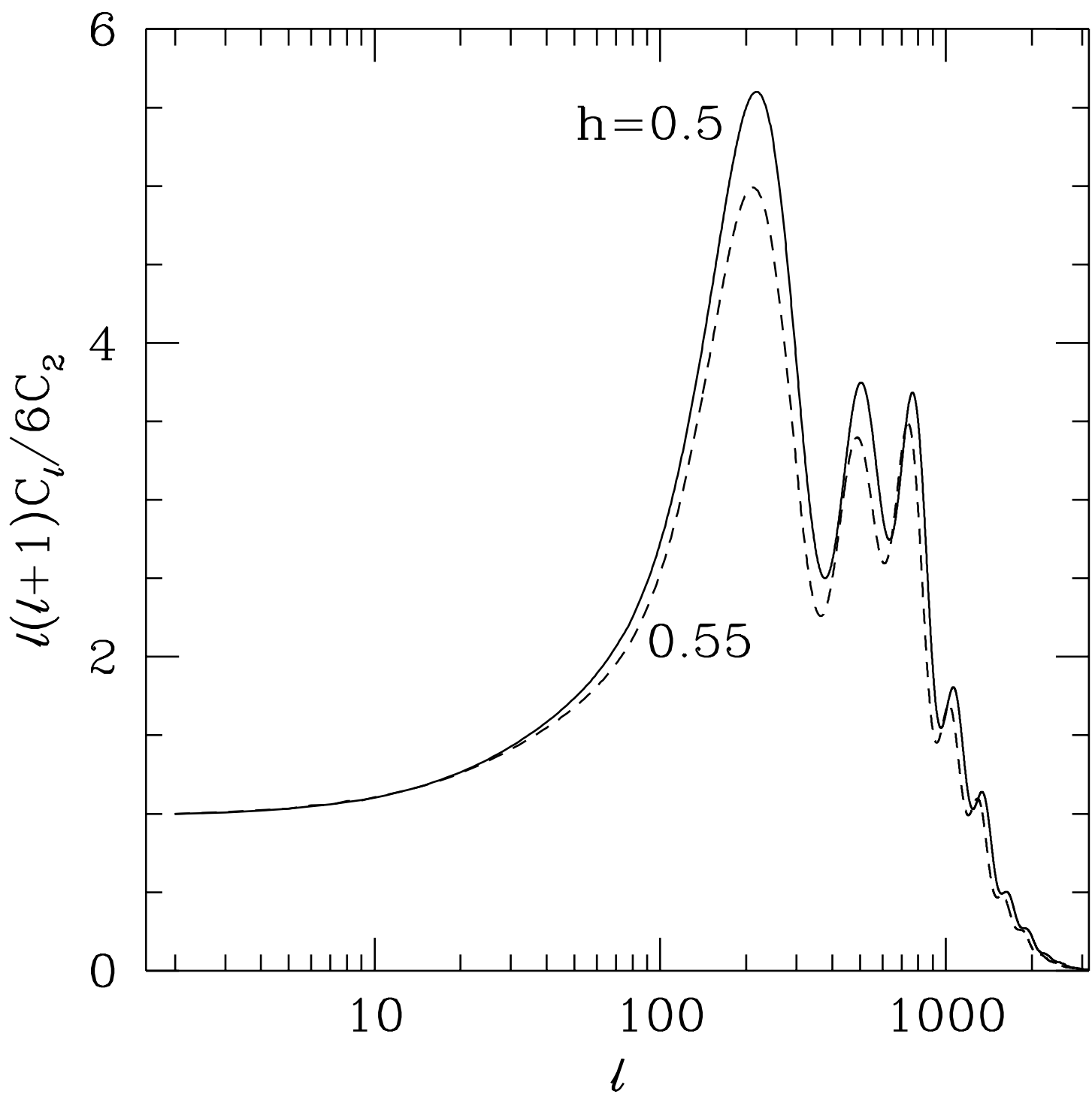

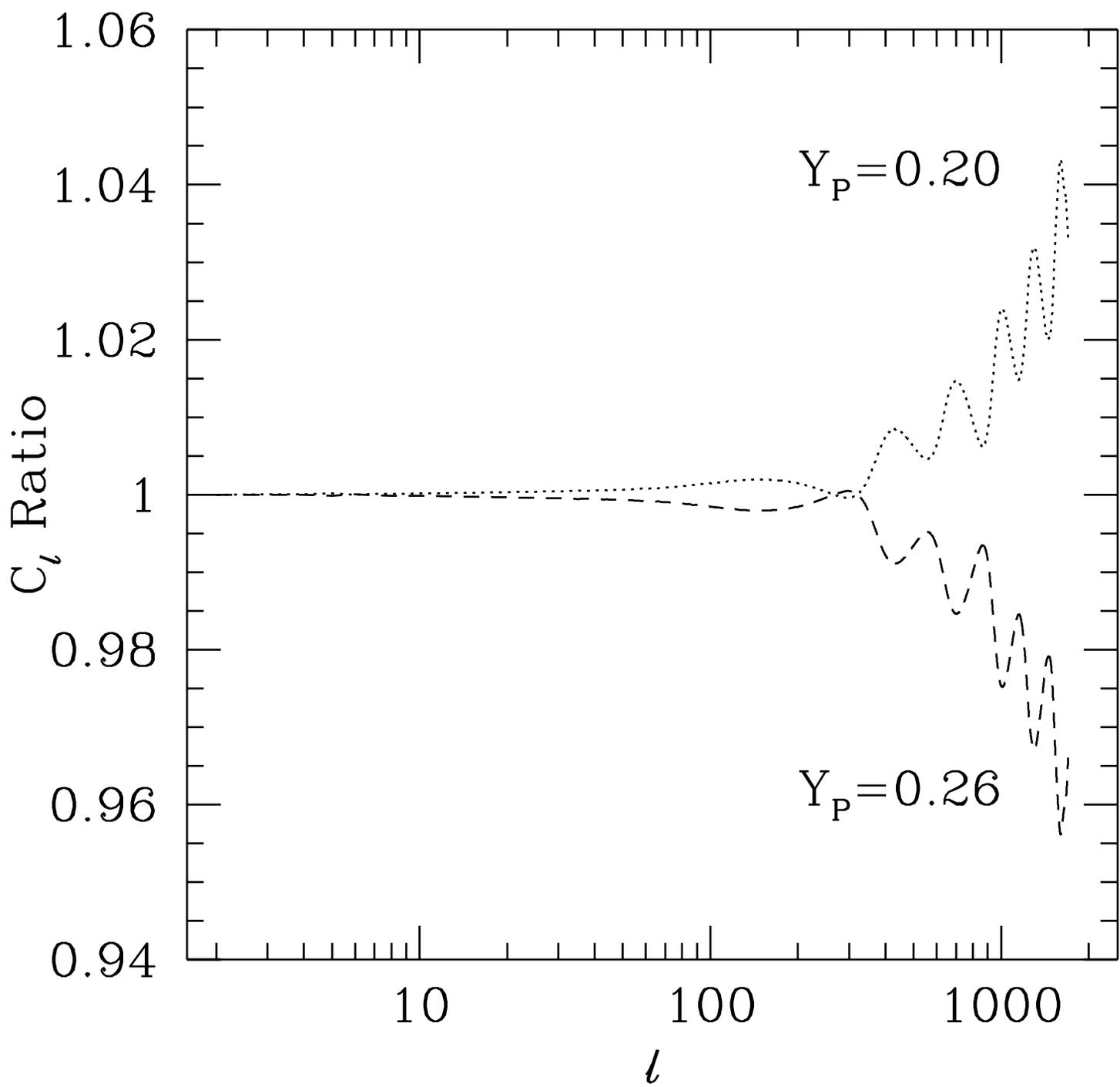

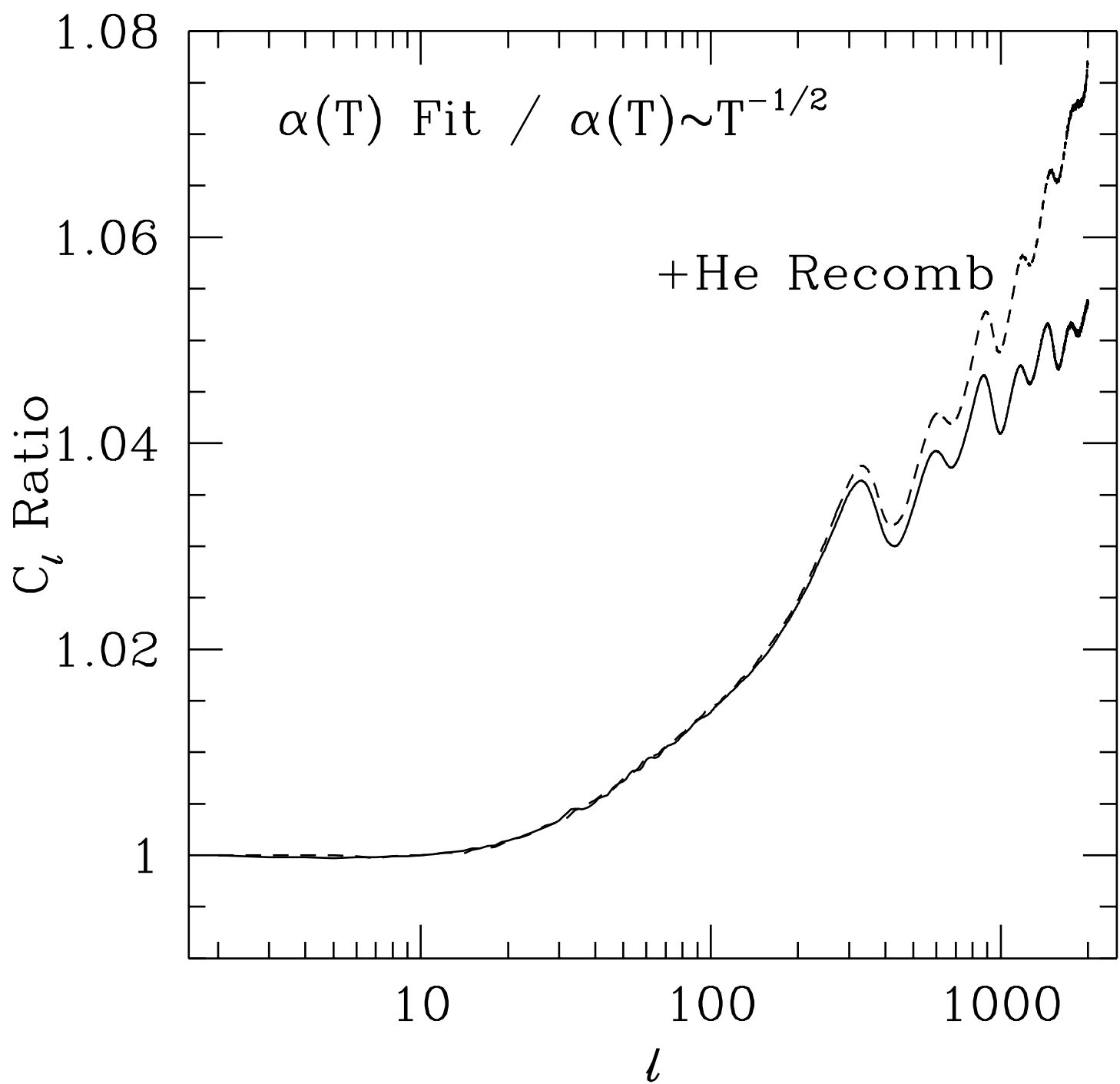

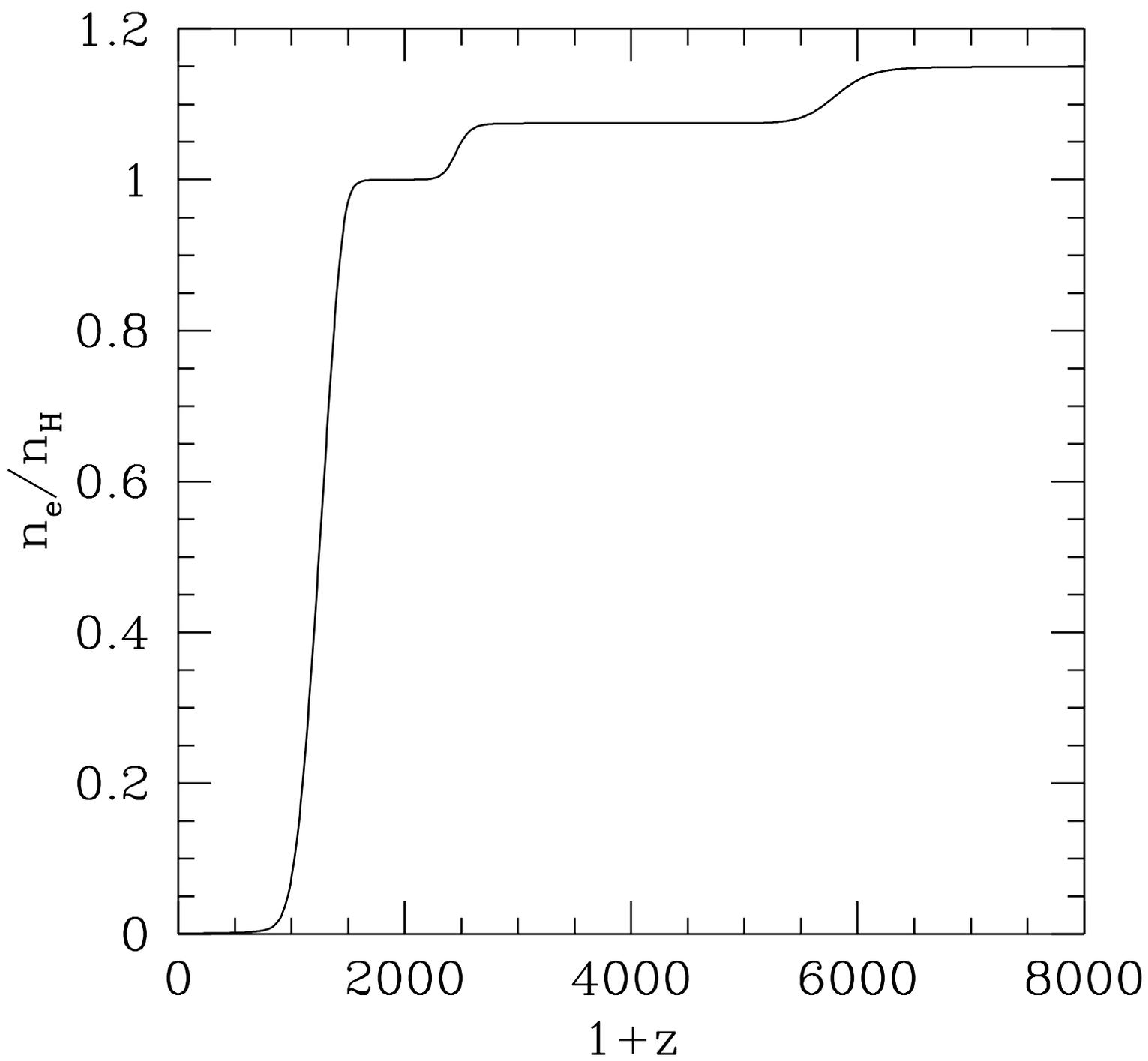

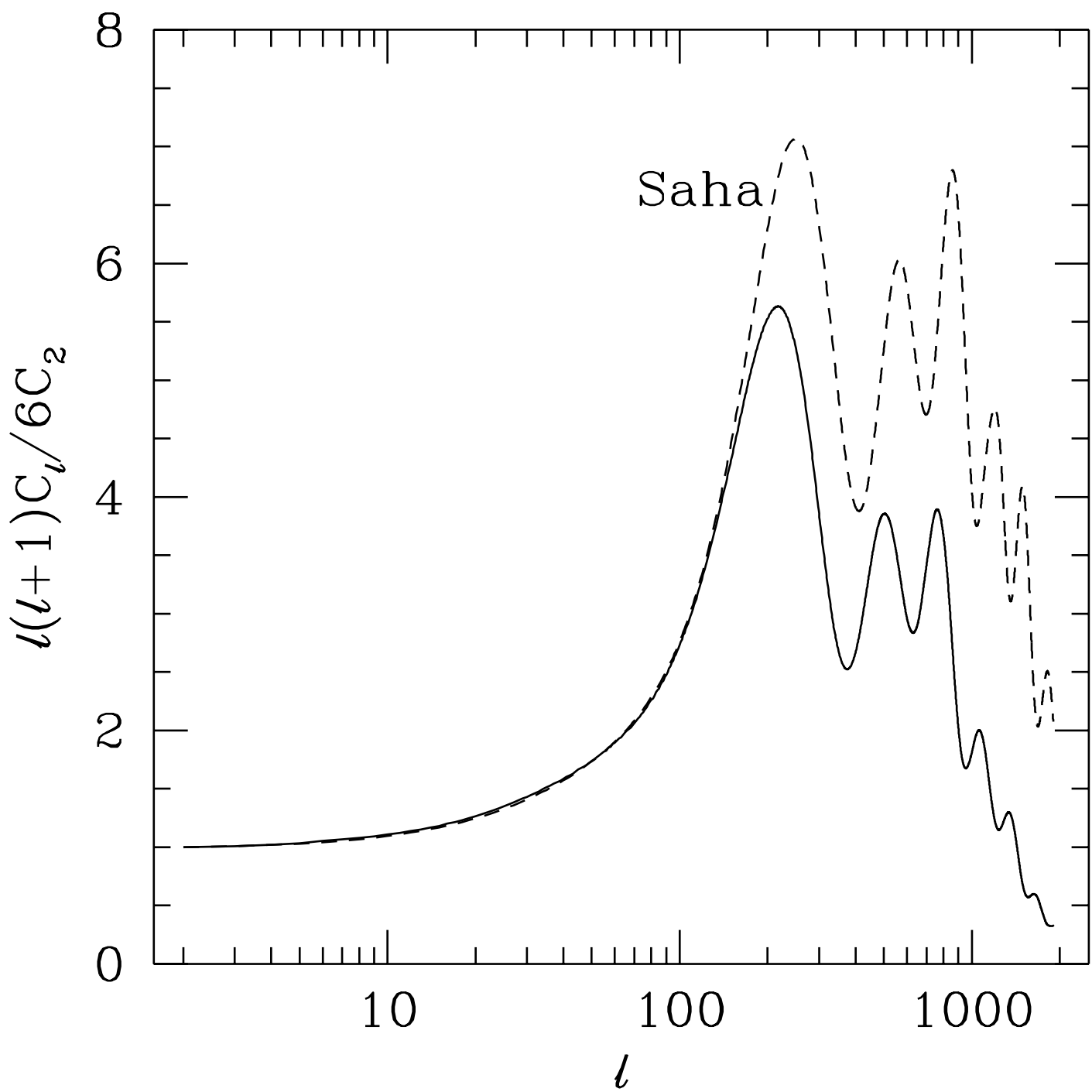

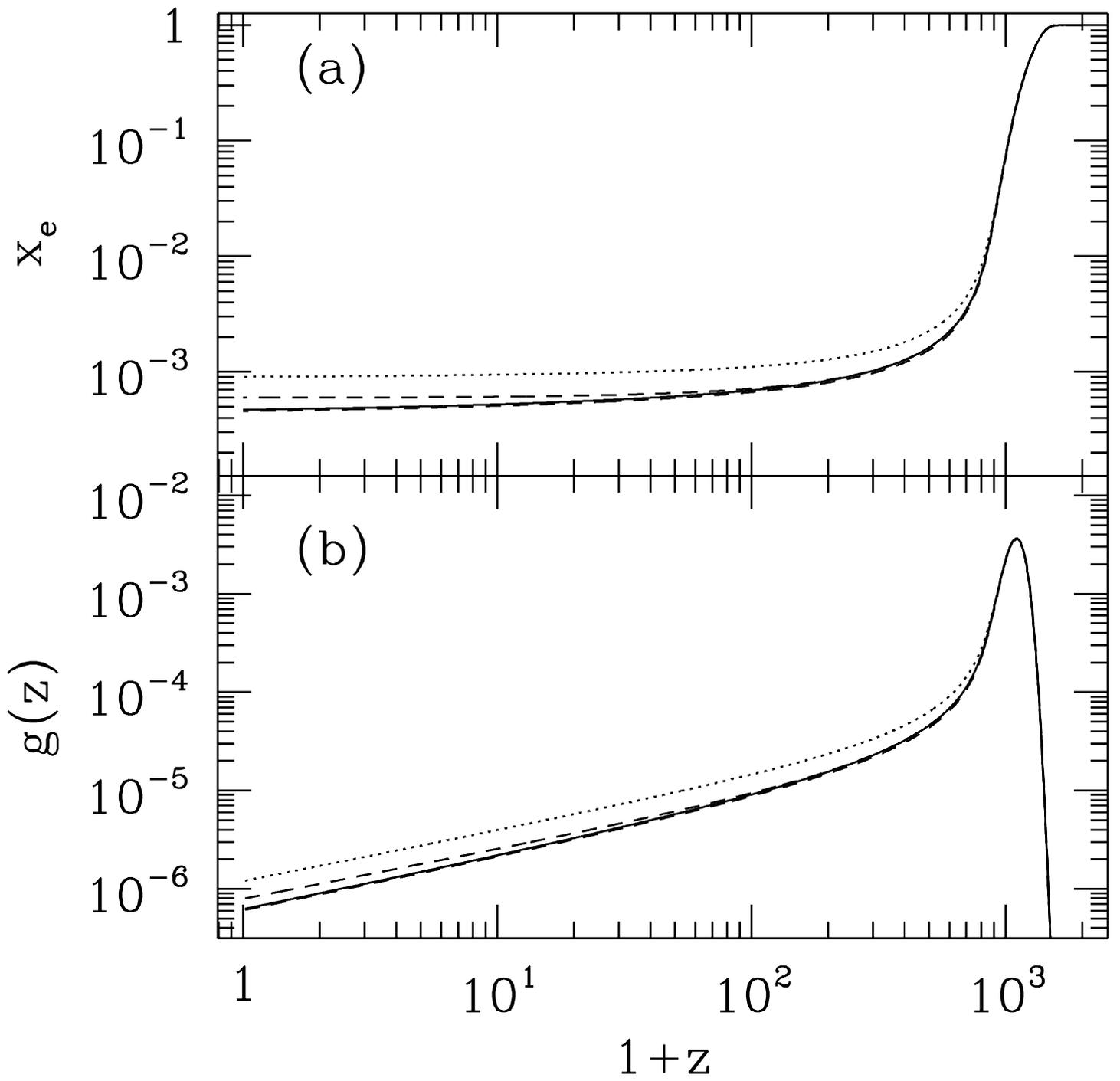

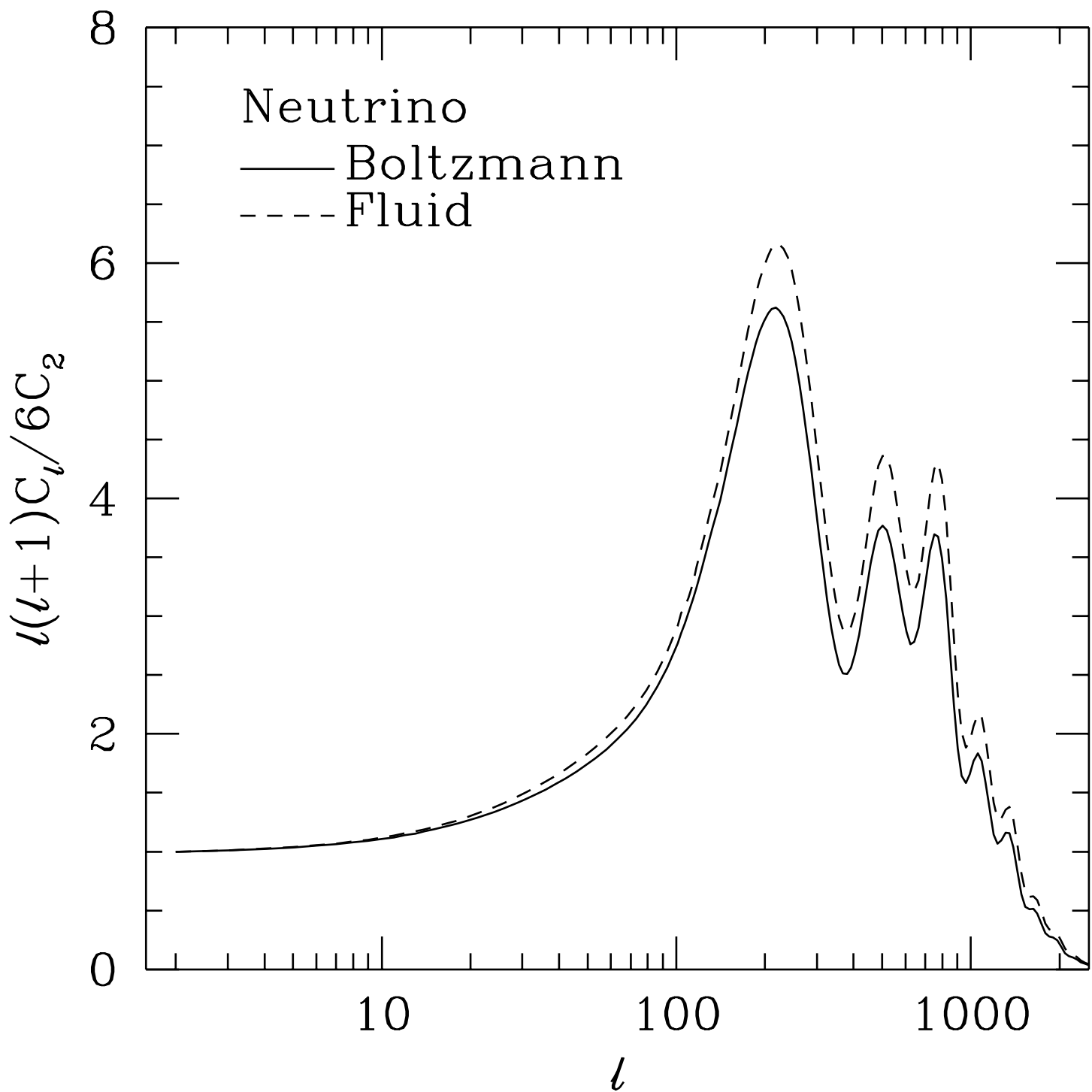

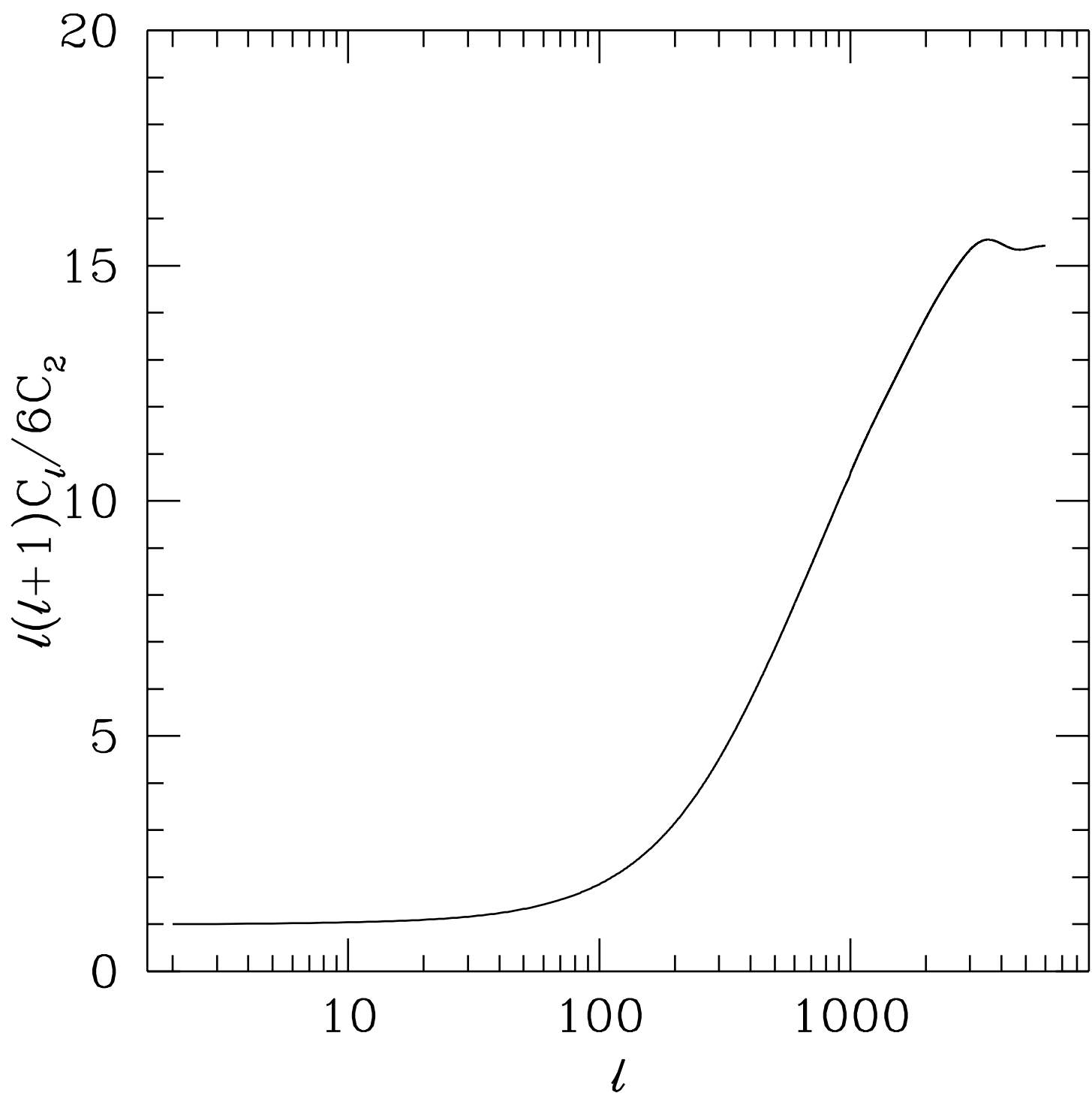

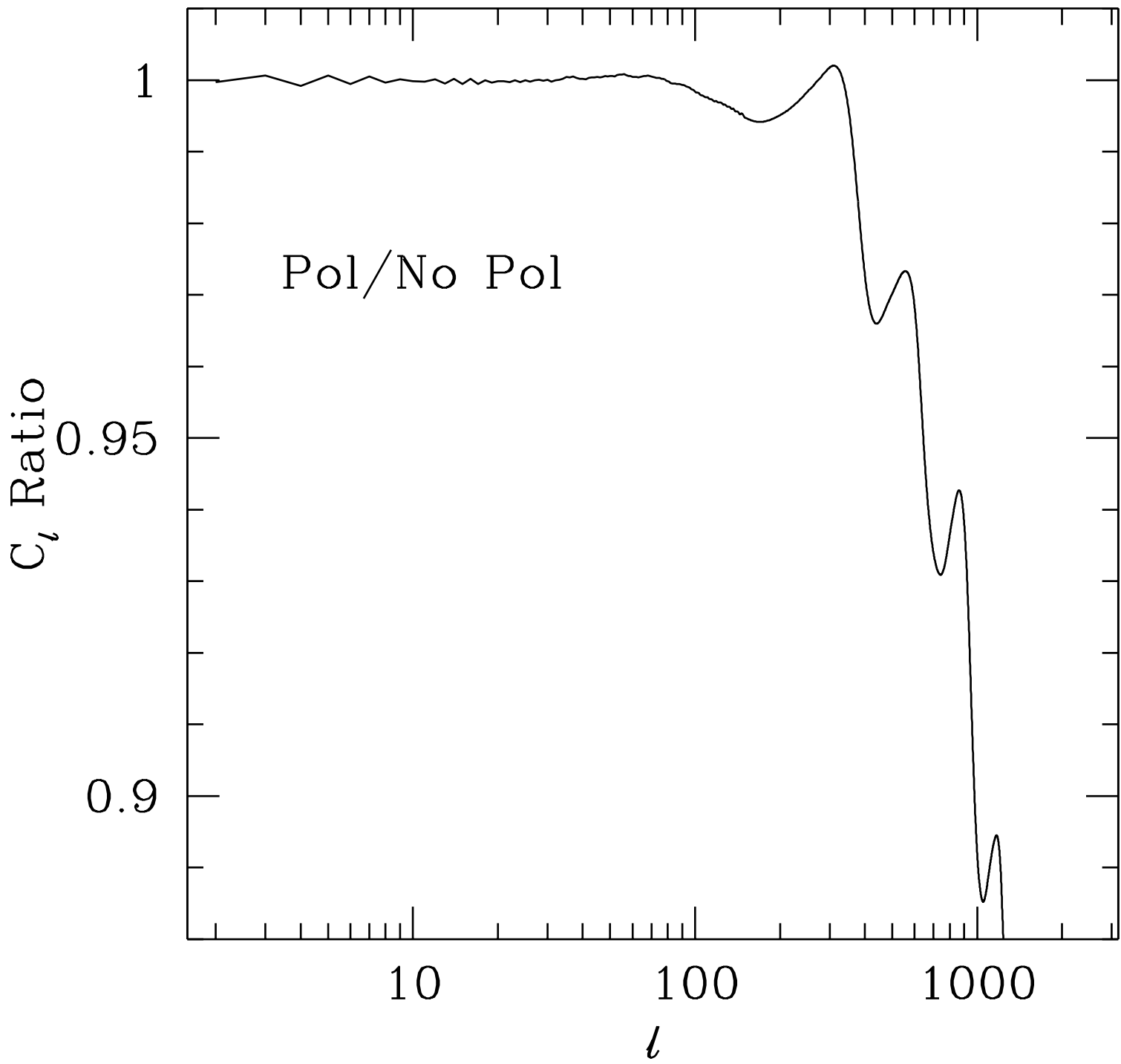

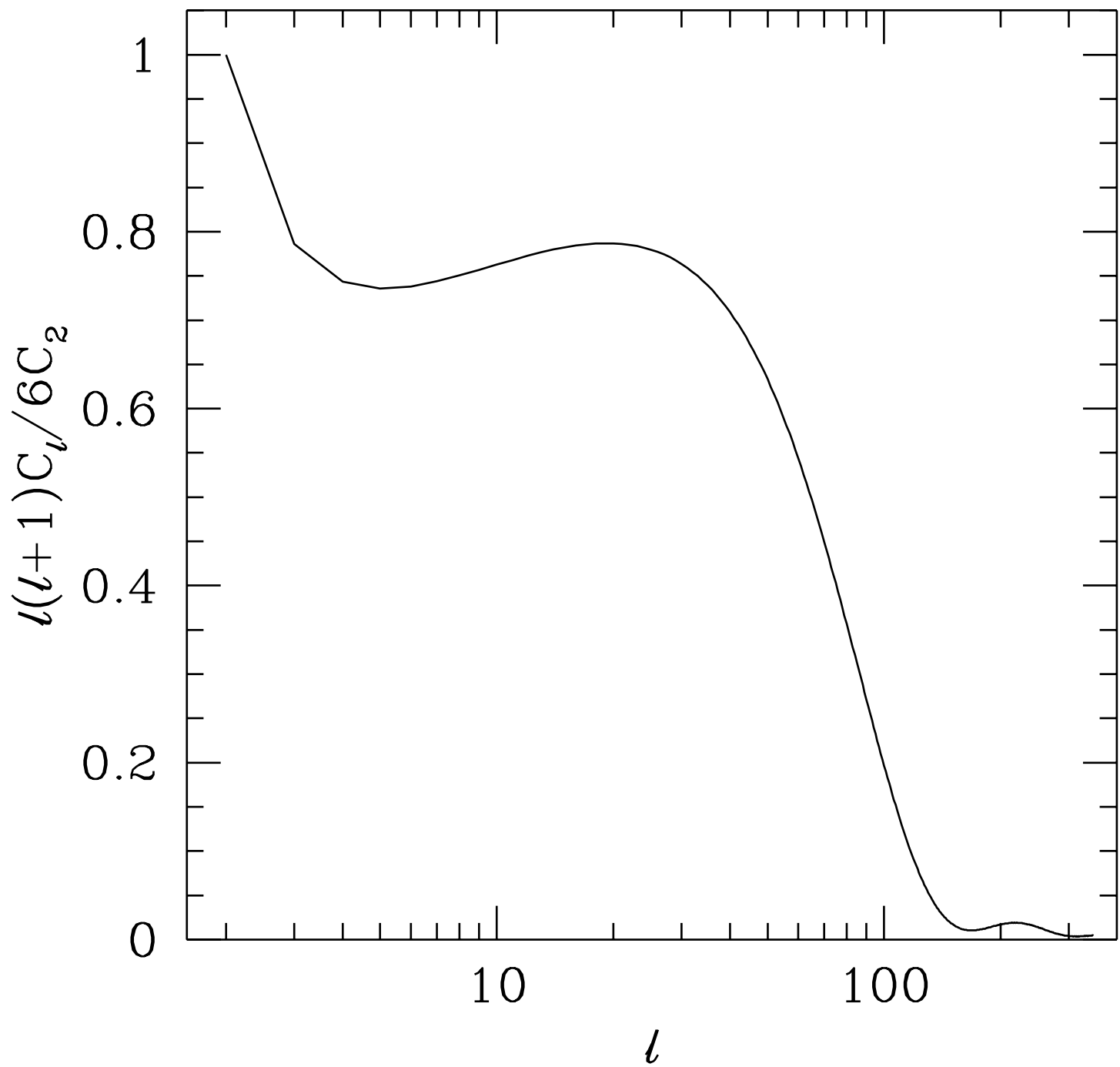

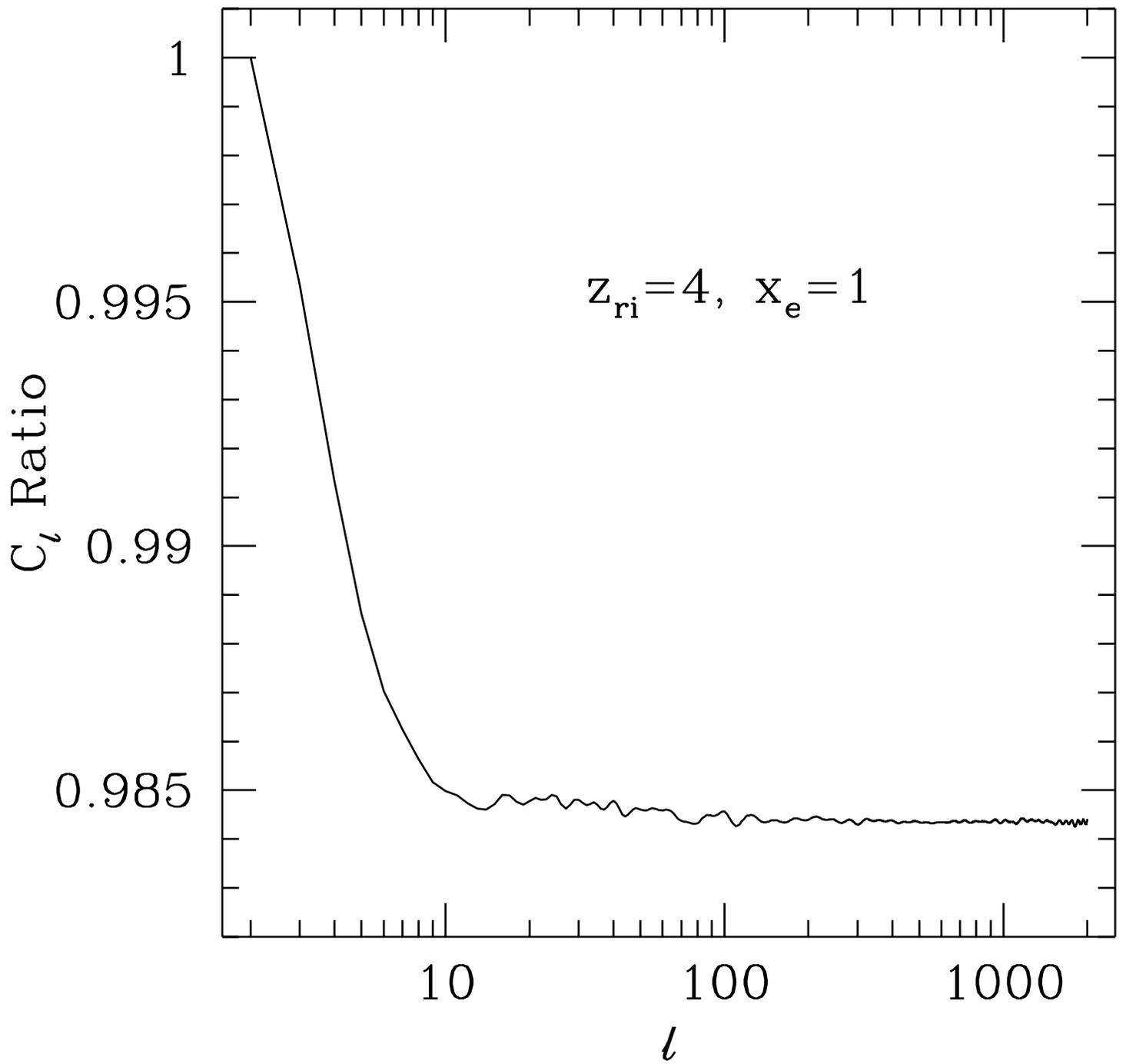

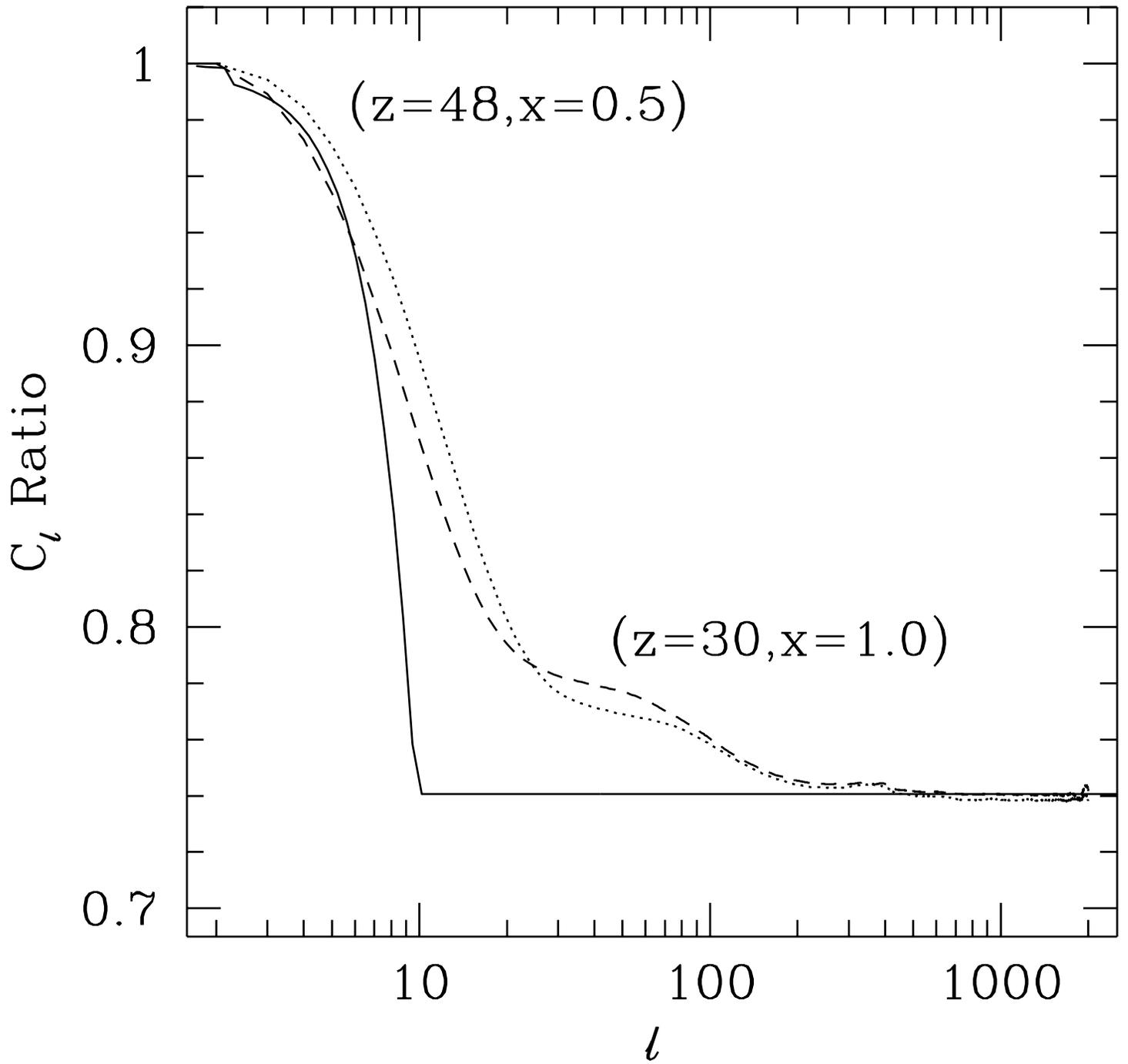

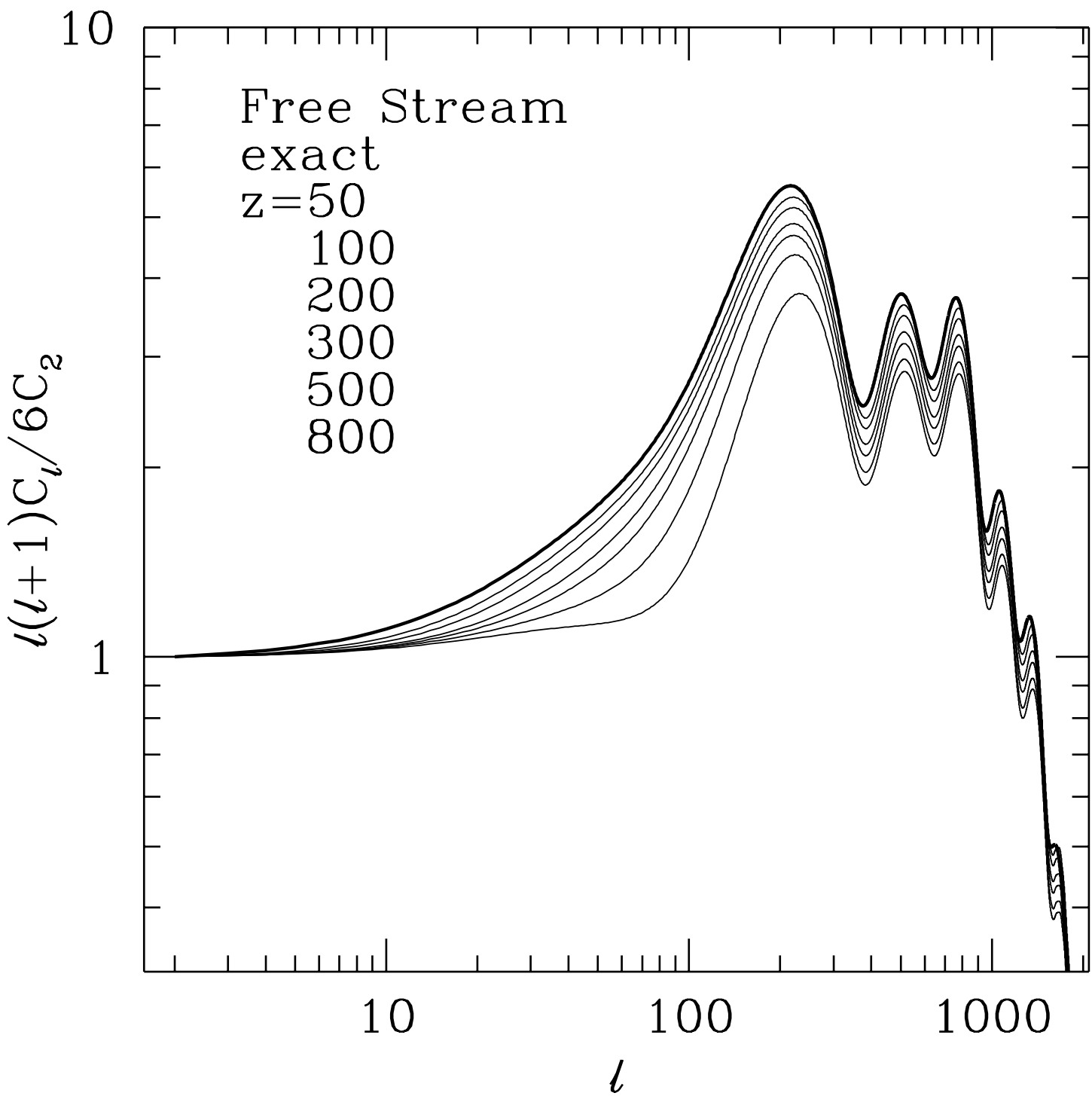

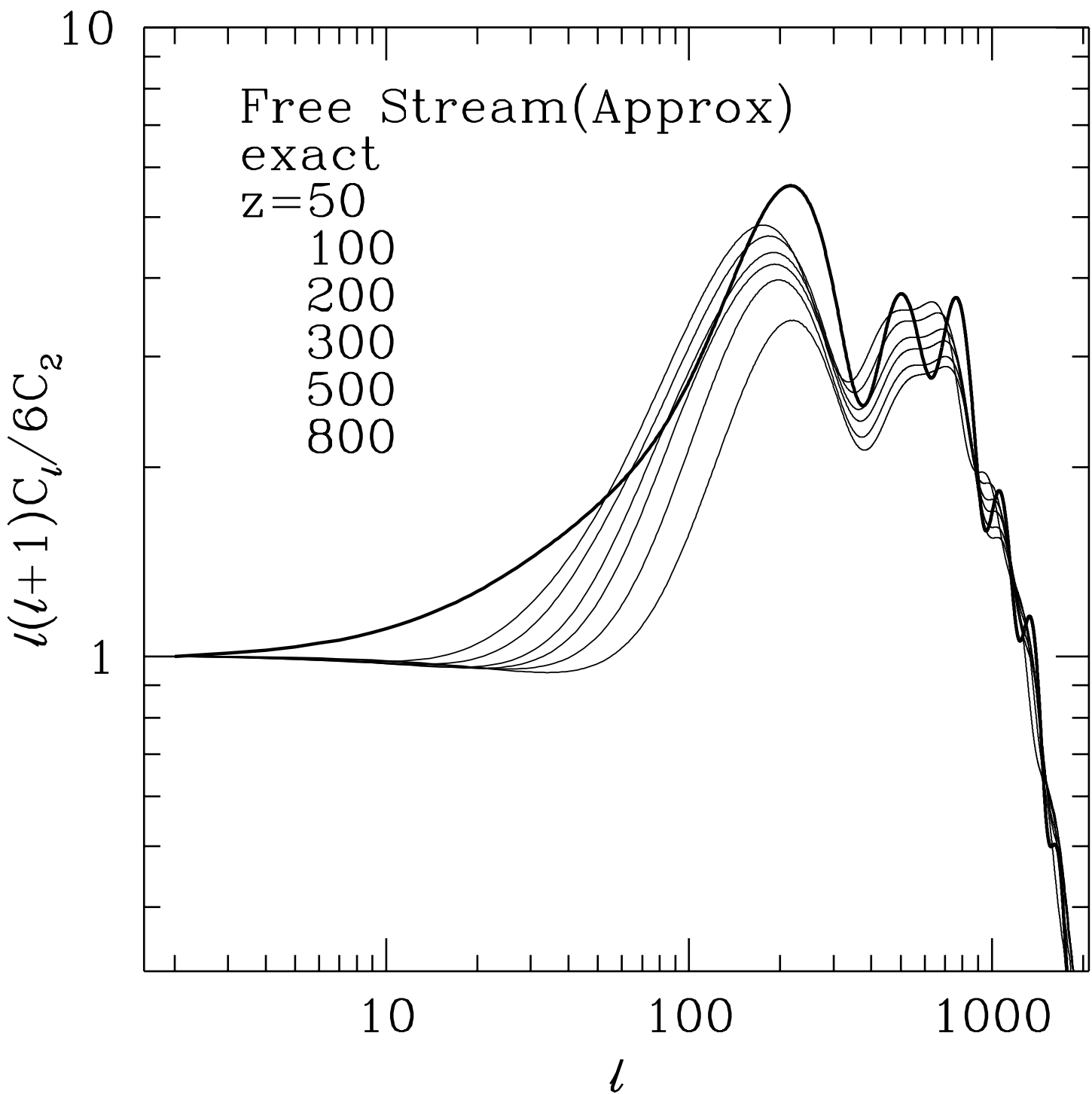

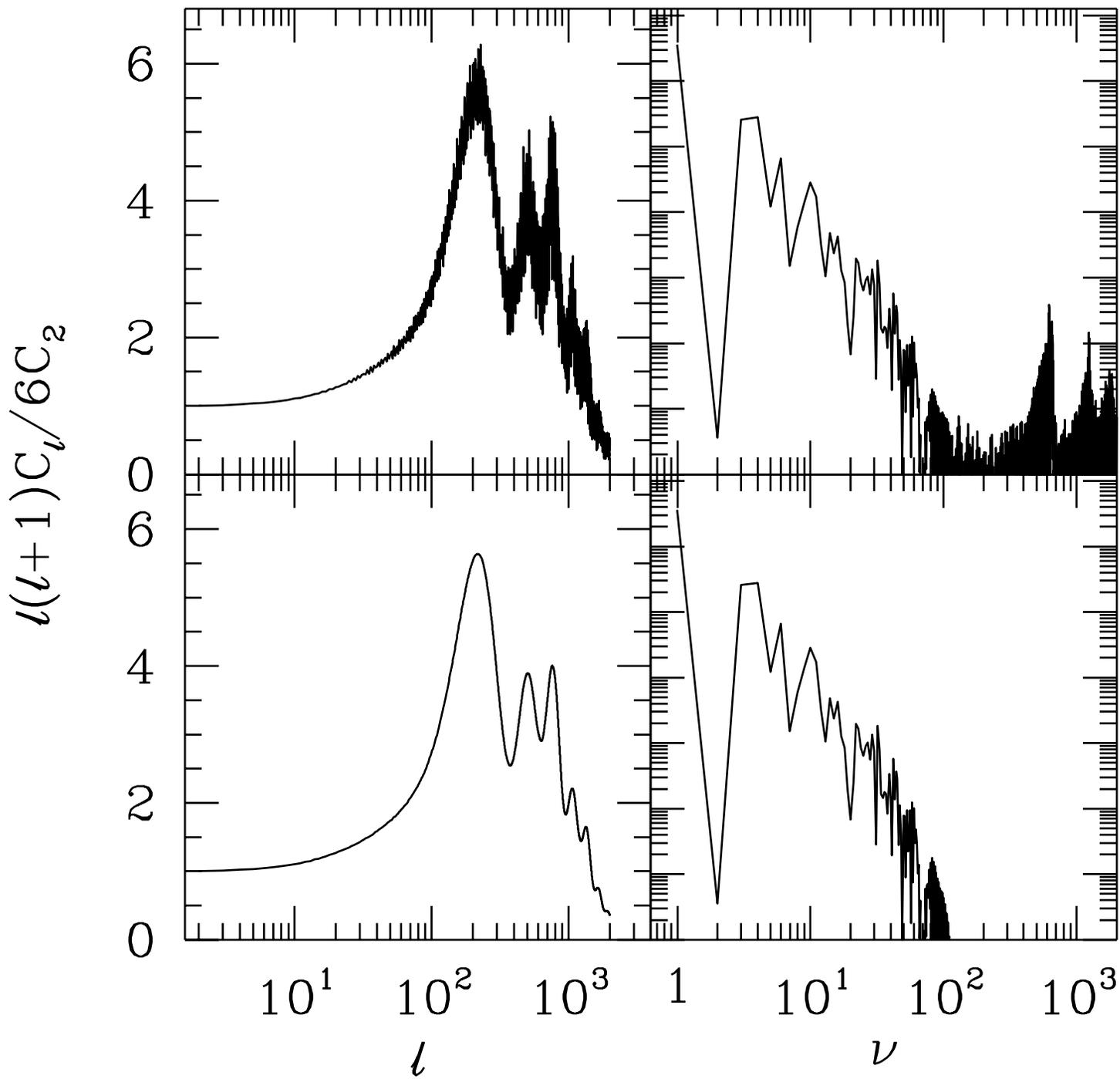

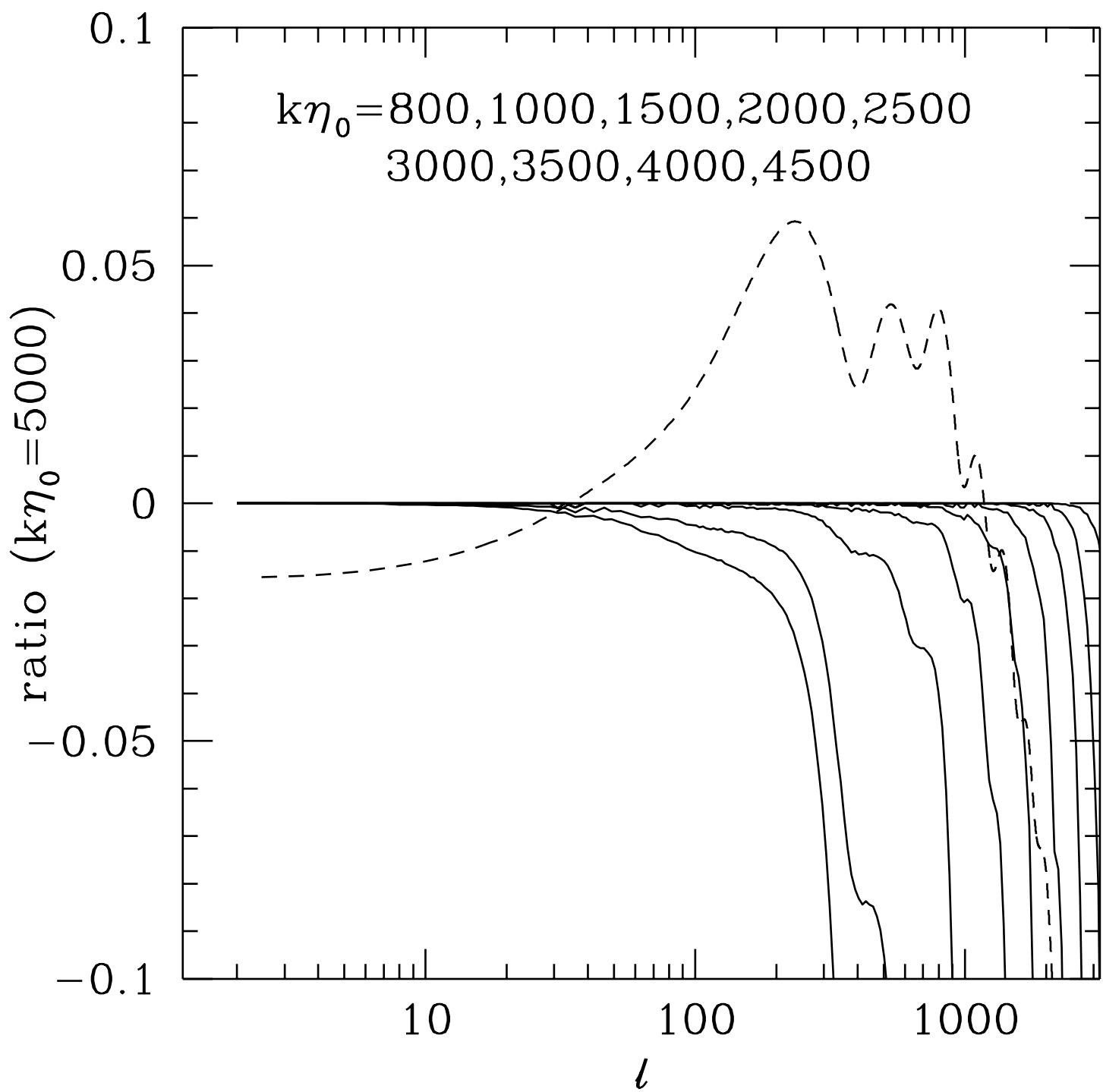

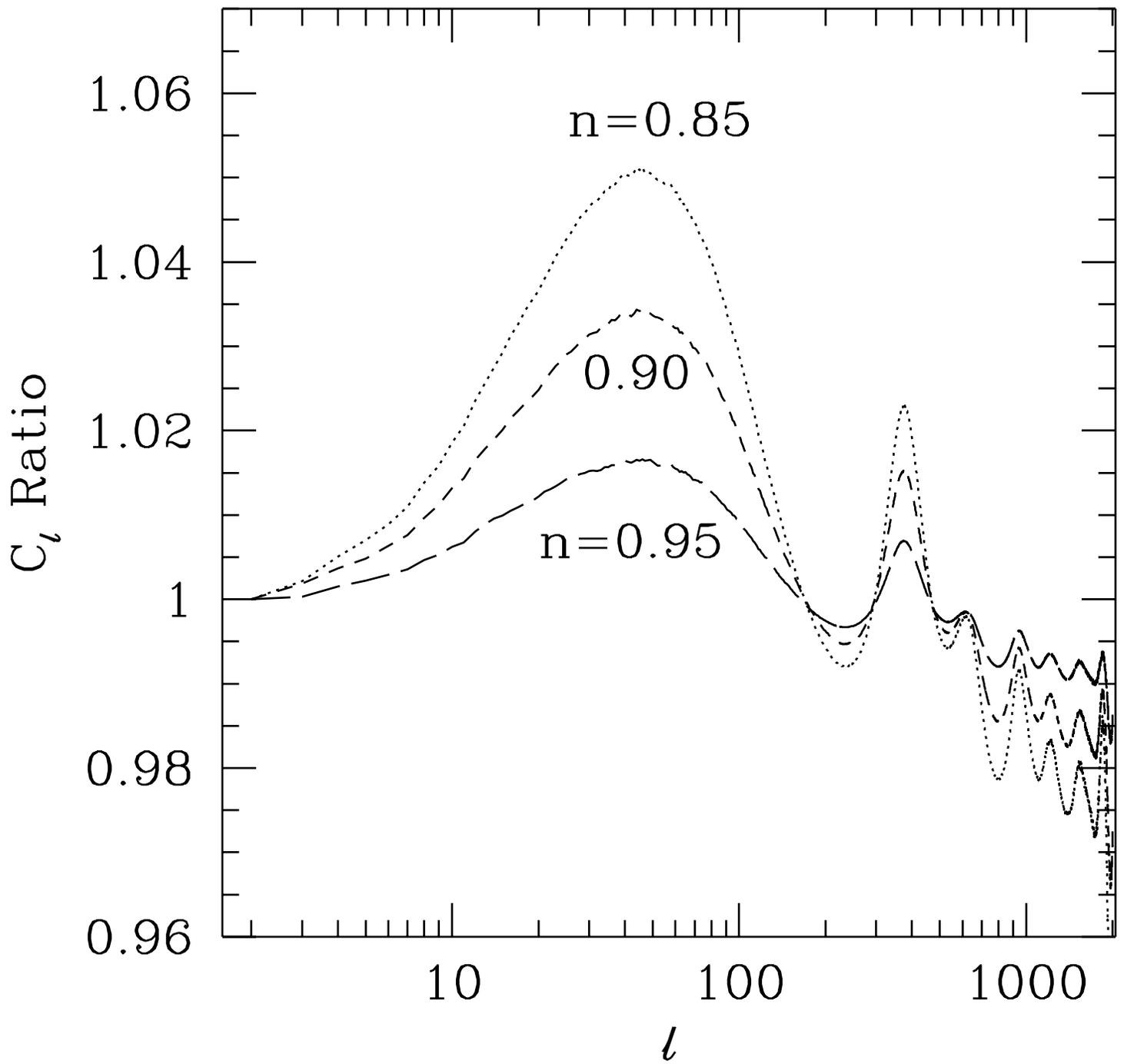